\documentclass[reprint,aps,floatfix,
  prd,
  superscriptaddress,
  nofootinbib
]{revtex4-2}
\usepackage[english]{babel}
\usepackage{multirow}
\usepackage{graphicx}
\usepackage{dcolumn}
\usepackage{amsmath}
\usepackage{amssymb}
\usepackage{hyperref}
\usepackage{gensymb}
\usepackage{xcolor}
\usepackage{caption}
\usepackage{subcaption}
\usepackage{enumitem}
\usepackage{soul}
\usepackage[normalem]{ulem}
\hypersetup{    
  colorlinks      = true,
  linkcolor       = {blue},
  linkbordercolor = {white},
  citecolor       = {blue},
  citebordercolor = {white},
  urlcolor        = {blue},
  urlbordercolor  = {white},
}
\usepackage{bm}
\usepackage{color}

\newcommand{\ben}{\begin{enumerate}}
\newcommand{\een}{\end{enumerate}}
\newcommand{\bea}{\begin{eqnarray}}
\newcommand{\eea}{\end{eqnarray}}
\newcommand{\be}{\begin{equation}}
\newcommand{\ee}{\end{equation}}

\def\rm{\rm}

\usepackage{scrextend}
\usepackage{comment}
\usepackage{booktabs}
\usepackage{graphicx}
\usepackage{slashed}

\usepackage{tikz}
\usetikzlibrary{positioning, shapes.geometric, arrows.meta}

\definecolor{ABcolor}{rgb}{1,0,0.8}

\def\nn{\nonumber}

\def\l{\left}
\def\r{\right}

\def\l{\left}
\def\r{\right}
\def\a{\alpha}
\def\b{\beta}

\def\d{\delta}

\def\w{\omega}

\def\La{\Lambda}

\def\O{\Omega}

\def\rm{\mathrm}

\def\*{\star}

\newcommand{\viz}{\text{viz.~}}
\newcommand{\ie}{\text{i.e.,~}}

\begin{document}

\renewcommand{\arraystretch}{1.5} 
\setlength{\tabcolsep}{0.2cm} 
\hfill{LITP-26-16}

\title{Transient Early Dark Energy–Like Dynamics as a Mechanism for Enhanced Early Structure Formation in the JWST Era}

\author{Abhik Bhattacharjee}
\email{abhikbhattacharjee@hri.res.in}
\address{\footnotesize Harish-Chandra Research Institute, Chhatnag Road, Jhunsi,
Prayagraj, Uttar Pradesh 211019, India}
\address{\footnotesize Homi Bhabha National Institute, Training School Complex, Anushakti Nagar, Mumbai
400094, India}

\author{Amlan Chakraborty}
\email{amlan.chakraborty@iiap.res.in}
\address{\footnotesize Indian Institute of Astrophysics, Bengaluru, Karnataka 560034, India} 

\author{Subinoy Das}
\email{subinoy@iiap.res.in}
\address{\footnotesize Indian Institute of Astrophysics, Bengaluru, Karnataka 560034, India}

\author{Anshuman Maharana}
\email{anshumanmaharana@hri.res.in}
\address{\footnotesize Harish-Chandra Research Institute, Chhatnag Road, Jhunsi,
Prayagraj, Uttar Pradesh 211019, India}
\address{\footnotesize Homi Bhabha National Institute, Training School Complex, Anushakti Nagar, Mumbai
400094, India}
\address{\footnotesize Leinweber Institute for Theoretical Physics, Randall Laboratory of Physics,
University of Michigan, Ann Arbor
450 Church St, Ann Arbor, MI 48109-1040, USA}

\author{Priyank Parashari}
\email{ppriyank@usc.edu}
\address{Department of Physics \& Astronomy, University of Southern California, Los Angeles, CA, 90089, USA}

\begin{abstract}
The discovery of massive galaxies at redshifts $z\gtrsim10$ by the James Webb Space Telescope (JWST) has renewed interest in cosmological mechanisms capable of enhancing early structure formation while preserving the successful large-scale predictions of the standard $\Lambda$CDM model. We investigate a phenomenological scenario in which an exotic dark matter species can undergo a transient early dark energy-like phase during the radiation-dominated era ($10^{-7}\lesssim a\lesssim10^{-5}$) before reverting to pressureless cold dark matter. We utilize the generalized dark matter framework to model this species, which is restricted to a sub-percent fraction of the total dark matter component by CMB, BAO and Type Ia supernova data. Its background and perturbation dynamics are characterized by a time-dependent equation of state, $w(a)$, and a time- as well as scale-dependent sound speed, $c_s^2(a,k)$. The temporary negative equation of state, combined with our phenomenological pressure-response prescription, induces a finite interval of negative effective sound speed squared. This triggers an instability-driven growth of density perturbations over a limited range of comoving scales, thereby enhancing the formation of early dark matter halos. We find that the enhanced halo abundance can substantially reduce the star-formation efficiencies required to reproduce the observed abundance of JWST galaxies relative to the standard $\Lambda$CDM scenario, especially at higher redshifts. Our results demonstrate that transient early dark energy-like dynamics in a subdominant dark matter component provide a viable mechanism for enhancing early structure formation and offer a new framework for interpreting the abundance of high-redshift galaxies observed by JWST and future surveys.

\end{abstract}

\maketitle
\section{Introduction} \label{sec:intro}

The James Webb Space Telescope (JWST) has opened an unprecedented observational window into the first few hundred million years of cosmic history, enabling the detection of galaxies at redshifts well beyond those previously accessible~\cite{Naidu_2022,Harikane_2023,curtislake2023spectroscopicconfirmationmetalpoorgalaxies,Fujimoto_2023,Carniani_2024}. These observations provide a unique probe of the earliest stages of galaxy formation and the growth of cosmic structure. Early analyses of JWST observations reported a population of unexpectedly massive galaxies at $z\gtrsim7$, leading to claims of significant tension with the standard Lambda Cold Dark Matter ($\Lambda$CDM) cosmology~\cite{Haslbauer_2022, Lovell_2022, Labb__2023,Boylan_Kolchin_2023, Santini_2023}. Subsequent spectroscopic observations, improved stellar-mass estimates, and more detailed galaxy-formation modeling have substantially alleviated the original discrepancy~\cite{finkelstein2023completeceersearlyuniverse,Arrabal_Haro_2023,Fujimoto_2023,barro2023extremelyredgalaxiesz59,Napolitano_2025}. Nevertheless, recent discoveries of galaxies at $z\gtrsim10$, including spectroscopically confirmed systems at $z\sim14$, continue to test the limits of current theoretical predictions~\cite{donnan2026spectroscopicconfirmationlargeluminous,wu2025jadesgsz141compactfaintgalaxy,naidu2026cosmicmiracleremarkablyluminous, harikane2023purespectroscopicconstraintsuv, Wang__2023, curtislake2023spectroscopicconfirmationmetalpoorgalaxies, harikane2024jwstalmakeckspectroscopic, robertsborsani2024extremesjwstspectroscopicbenchmark}. Although current observations no longer indicate a severe crisis for $\Lambda$CDM, the abundance of galaxies during cosmic dawn remains a powerful probe of the growth of structure and the underlying properties of dark matter. In particular, cosmological scenarios that enhance the abundance of early dark matter (DM) halos can substantially reduce the star-formation efficiencies required to reproduce the observed ultraviolet luminosity functions.

Several astrophysical mechanisms and galaxy-formation scenarios have been proposed to explain the observed abundance of luminous galaxies at $z\gtrsim10$ within the standard $\Lambda$CDM framework~\cite{harikane2023purespectroscopicconstraintsuv, driskell2024populationsynthesisastrophysicalinference}. These include enhanced star-formation efficiencies in compact, dense, and low-metallicity galaxies at high redshift~\cite{Harikane_2023,Fukushima_2021,Dekel_2023}, a non-negligible contribution of active galactic nuclei (AGN) to the observed ultraviolet luminosities~\cite{Bunker_2023,Larson_2023,harikane2023jwstnirspeccensusbroadlineagns}, a top-heavy initial mass function (IMF) that increases the UV output per unit star-formation rate~\cite{Chon_2022,Steinhardt_2023,Harikane_2023}, and a large intrinsic scatter in the halo mass -- star-formation rate ($M_h-\rm{SFR}$) relation arising from bursty star-formation histories~\cite{Mason_2015,Harikane_2016,Harikane_2023,Sarkar_2026,lazare2026galaxiesburstiiimplications}.

An alternative possibility is that the observed abundance of early galaxies reflects a genuine modification of the growth of density perturbations rather than the efficiency of baryonic star-formation and other astrophysical processes. Accordingly, a number of cosmological scenarios have been explored in which the abundance of early galaxies is increased through modifications to the growth of structure. These include revised halo-collapse prescriptions and halo mass functions~\cite{Fakhry_2026}, the seeding of early structure by topological defects such as cosmic strings~\cite{Jiao_2023,blamart2025uvluminosityfunctionshst}, modifications to the primordial power spectrum through a blue tilt~\cite{Parashari_2023}, running~\cite{kobayashi2026bluetiltedrunningsjwstearly}, non-Gaussianity~\cite{Biagetti_2023}, or localized enhancements in the matter power spectrum~\cite{Tkachev_2023, sabti2024insightshstultramassivegalaxies}, alternative DM scenarios involving fuzzy dark matter~\cite{Gong_2023, ghara2026fuzzydarkmatterhalo}, axion-like particles~\cite{roy2023sensitivityjwstevscaledecaying,Bird_2024}, axion miniclusters~\cite{H_tsi_2023}, or primordial black holes~\cite{Liu_2022,H_tsi_2023}, as well as departures from the standard expansion history through early dark energy, dynamical dark energy, or a negative cosmological constant~\cite{Forconi_2024,Adil_2023,Menci_2024,menci2024excessjwstbrightgalaxies,shen2024earlygalaxiesearlydark}. Such scenarios can increase the abundance of early DM halos and thereby reduce the star-formation efficiencies required to reproduce the observed ultraviolet luminosities. 

Despite the diversity of proposed cosmological solutions, many rely on changes to the background expansion history over an extended cosmological epoch or fundamental modifications to the primordial power spectrum. An attractive alternative is to transiently modify the dynamics of only a sub-component of the dark matter, thereby enhancing the growth of density perturbations over a restricted range of scales while leaving both the primordial spectrum and the successful large-scale evolution of the standard $\Lambda$CDM cosmology essentially unchanged. Such a mechanism naturally increases the abundance of early DM halos without requiring prolonged departures from the standard background evolution. This forms the central idea explored in the present work.

We consider a phenomenological realization within the Generalized Dark Matter (GDM) framework~\cite{Hu_1998}. Specifically, we study a subdominant DM component whose equation of state parameter undergoes a temporary departure from its CDM value during the radiation-dominated (RD) era. This can give rise to a transient Early Dark Energy (EDE)-like phase, in the sense that the equation of state parameter of the GDM fluid becomes negative and can remain below $-1/3$ for a finite interval. 
While the transient behavior considered here is not derived from a specific microscopic model, it is partly motivated by coupled dark sector scenarios involving an interacting scalar field $\phi$ and a DM component $\psi$, where negative effective sound speed squared resulting from the negative equation of state parameter of the coupled fluid can lead to instabilities and enhanced growth of density perturbations, as explored in Refs.~\cite{Gogoi_2021,Chakraborty_2022}. These models are themselves motivated by mass-varying dark matter/neutrino scenarios, in which the mass of the matter species depends on a dynamical scalar field and the coupled system can exhibit instability-driven clustering (see, for example, Refs.~\cite{Afshordi_2005,Bjaelde_2009}). 

Phenomenological departures from the standard pressureless DM equation of state have also recently been considered in Ref.~\cite{braglia2025exoticdarkmatterdesi} to study the late-time DESI anomaly. Moreover, transient evolution of the equation of state parameter has also been realized in the sterile neutrino-pseudoscalar scenario of Ref.~\cite{sharma2026recoupleddarkradiationreconciling}, where the effective equation of state temporarily departs from its asymptotic radiation-like value. Although the asymptotic behavior in the present work is matter-like rather than radiation-like, this example demonstrates that transient, non-monotonic evolution of the equation of state parameter can arise naturally in explicit dark sector models. Our objective is therefore not to construct a complete microscopic model\footnote{However, see Appendix~\ref{sec:micro}.}, but rather to identify the cosmological signatures of a transient EDE-like phase and assess whether such transient dark-sector dynamics can alleviate the unusually high star-formation efficiencies required within the standard $\Lambda$CDM framework~\cite{Yung_2025}.

The remainder of this paper is organized as follows. In Sec.~\ref{sec:model}, we introduce the GDM model and discuss its background and perturbation evolution. In Sec.~\ref{sec:Pk_UVLF}, we outline the cosmological observables used in our analysis, including the matter power spectrum, halo mass function, and ultraviolet luminosity function. Our results are presented in Sec.~\ref{sec:results}, and we summarize our conclusions in Sec.~\ref{sec:conclusion}.

\section{The GDM Model} \label{sec:model}
The GDM framework, introduced in Ref.~\cite{Hu_1998}, provides a phenomenological description of DM beyond the standard CDM paradigm. While CDM is modeled as a pressureless perfect fluid, a variety of well-motivated DM candidates can exhibit non-zero pressure, viscosity, or other imperfect fluid effects at the level of linear perturbations. Examples include warm dark matter \cite{Armendariz_Picon_2014, Piattella_2016}, massive neutrinos~\cite{PhysRevD.82.089901, Oldengott_2015, PhysRevD.87.103515}, interacting dark matter~\cite{PhysRevD.87.063509, PhysRevD.66.083505, PhysRevD.81.043507,  Wilkinson_2014_2, Wilkinson_2014, Amendola_2000, PhysRevD.88.083505}, etc. Rather than specifying a particular microphysical model, the GDM framework models DM as an imperfect fluid characterized by three parameters: an equation of state (EOS) parameter $w_g(a)=\bar P_g/\bar\rho_g$ ($\bar P_g$ and $\bar\rho_g$ being the background pressure and energy density of the fluid, respectively), the effective sound speed $c_s^2(a,k)$ and the viscosity parameter $c_\rm{vis}^2(a,k)$. While the EOS parameter is only time-dependent, the latter two can be time- as well as scale-dependent~\cite{Kopp_2016}. In the CDM limit, these parameters reduce to $w_g=c_s^2=c_\rm{vis}^2=0$.
\subsection{Background evolution}
We consider an exotic DM component $X$ which makes up a fraction $f_X$ of the total DM content of the Universe today. It is modeled as a GDM fluid with background EOS parameter $w_X(a)$ given by:
\begin{align}\label{eos}
    w_X(a) = w_p \exp\left[ -\frac{1}{2}\l(\frac{\ln{(a/a_p)}}{\sigma_w}\r)^2 \right]\quad(w_p<0).
\end{align}
The EOS therefore exhibits a transient Gaussian dip for a range of scale factor $a$, centered around $a_p$, while asymptotically approaching zero at early ($a\ll a_p$) and late ($a\gg a_p$) times. The depth and width (in $\ln{a}$) of the dip are controlled by the parameters $w_p$ and $\sigma_w$, respectively. Suppose the transient epoch occurs between scale factors $a_i$ and $a_f$ in the RD era. Specifically, we identify $a_i=a_pe^{-3\sigma_w}$ and $a_f=a_pe^{3\sigma_w}$, which then allows us to fix $a_p$ and $\sigma_w$ as follows:
\begin{align}
    a_p=\sqrt{a_ia_f}\quad\text{and}\quad \sigma_w=\frac{1}{6}\ln{\l(\frac{a_f}{a_i}\r)}.
\end{align}
The adiabatic sound speed is derived from the background evolution as
\begin{align}
    c_a^2=w_X-\frac{\dot w_X}{3\mathcal H(1+w_X)},
\end{align}
with the overdot denoting derivative with respect to the conformal time $\tau$ and $\mathcal H\equiv\dot a/a=aH$ is the conformal Hubble parameter. 

\begin{figure}[htbp]
    \centering
    \includegraphics[width=\linewidth]{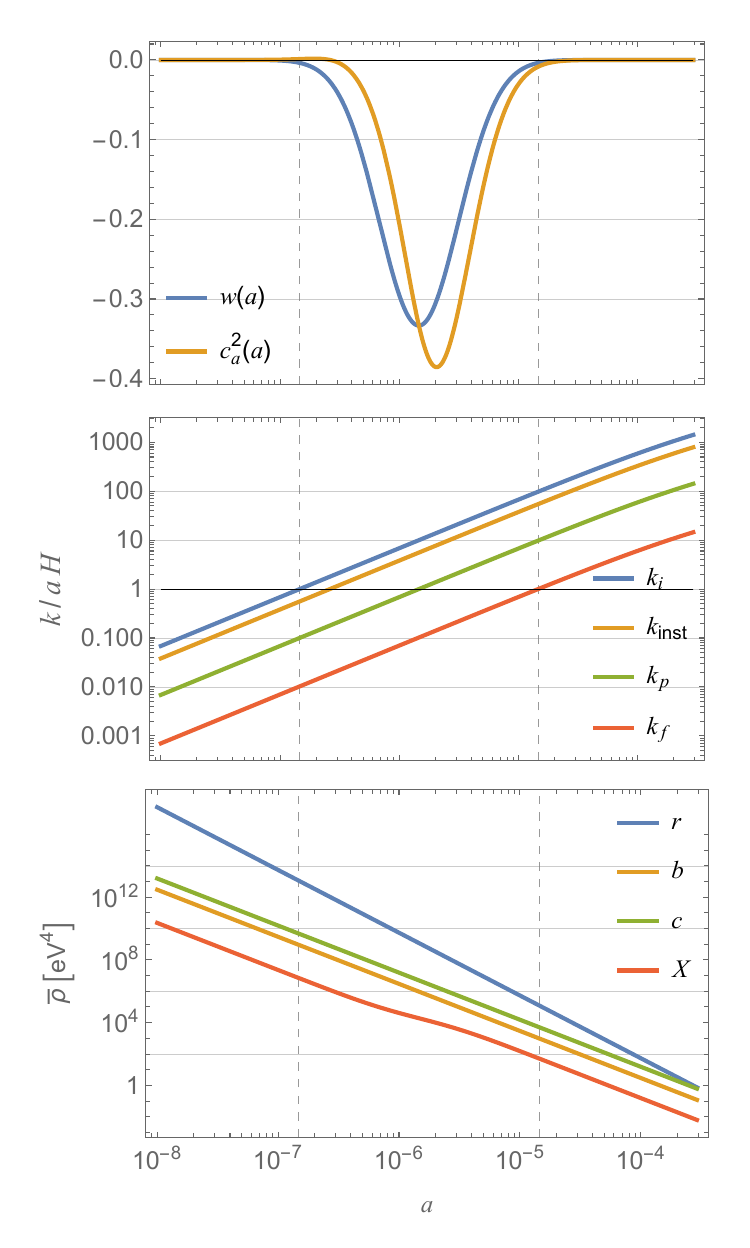}
    \caption{Evolution of the EOS parameter and the adiabatic sound speed with $w_p=-1/3$ (top), the ratio $k/aH$ for some representative modes illustrating their horizon-entry relative to the transient epoch (middle), and comparison of the background energy densities of radiation $(r)$, CDM $(c)$ and baryons $(b)$, and the exotic DM $X$ with $f_X=0.01$ (bottom). The vertical dashed lines correspond to $a_i$ and $a_f$. 
    }\label{fig:bg}
\end{figure}

For our numerical analysis, we set $a_i=1.46\times10^{-7}$ and $a_f=1.46\times10^{-5}$ throughout, which implies $a_p=1.46\times10^{-6}$ and $\sigma_w=0.767$. The background evolution of the model is illustrated in Fig.~\ref{fig:bg}. The top panel shows the transient Gaussian-dip in the EOS parameter, with $w_p=-1/3$, and the derived adiabatic sound speed. Note that the instability window, corresponding to negative $c_a^2$, begins at $a_\rm{inst}>a_i$. The middle panel displays the evolution of $k/aH$ for $k_i\approx28.15,~k_\rm{inst}\approx15.71,~k_p\approx2.82,$ and $k_f\approx 0.29~h/\rm{Mpc}$ corresponding to modes entering the horizon at $a_i,~a_\rm{inst},~a_p$, and $a_f$, respectively. The transient negative-$w_X(a)$ phase therefore spans the horizon-entry epoch of modes in the range $k\sim1-10~h/\rm{Mpc}$ relevant for JWST. Finally, the bottom panel compares the background energy densities of radiation, baryons, CDM, and the exotic DM component (X) with $f_X=0.01$. The vertical dashed lines mark the beginning and end of the transient.

\subsection{Linear Perturbations}
\begin{figure*}[htbp]
    \centering
    \begin{subfigure}{0.48\textwidth}
        \centering
        \includegraphics[width=\textwidth]{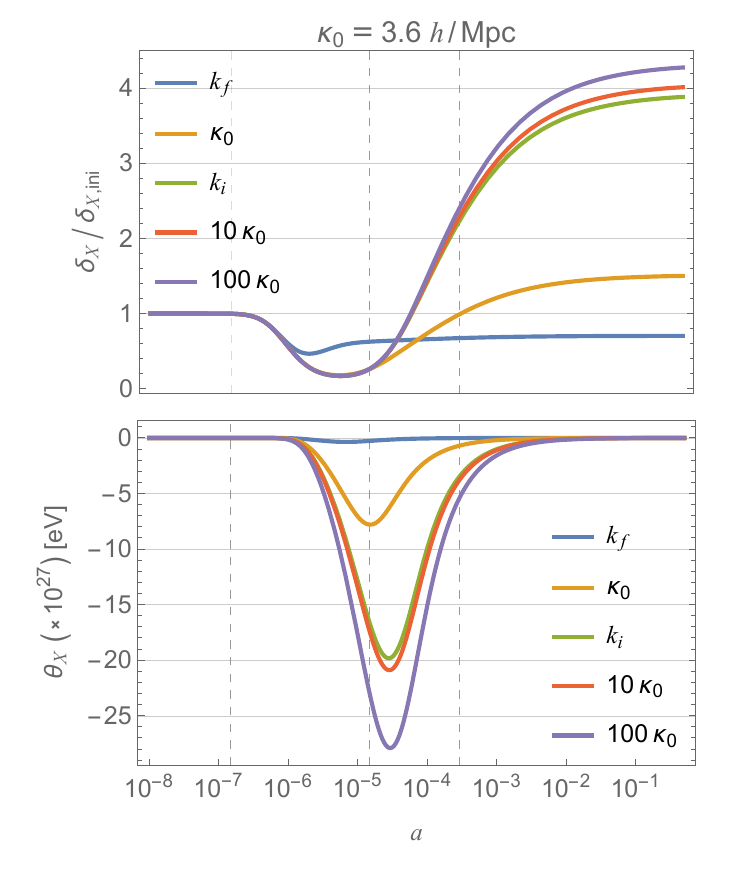}
    \end{subfigure}
    \hfill 
    \begin{subfigure}{0.48\textwidth}
        \centering
        \includegraphics[width=\textwidth]{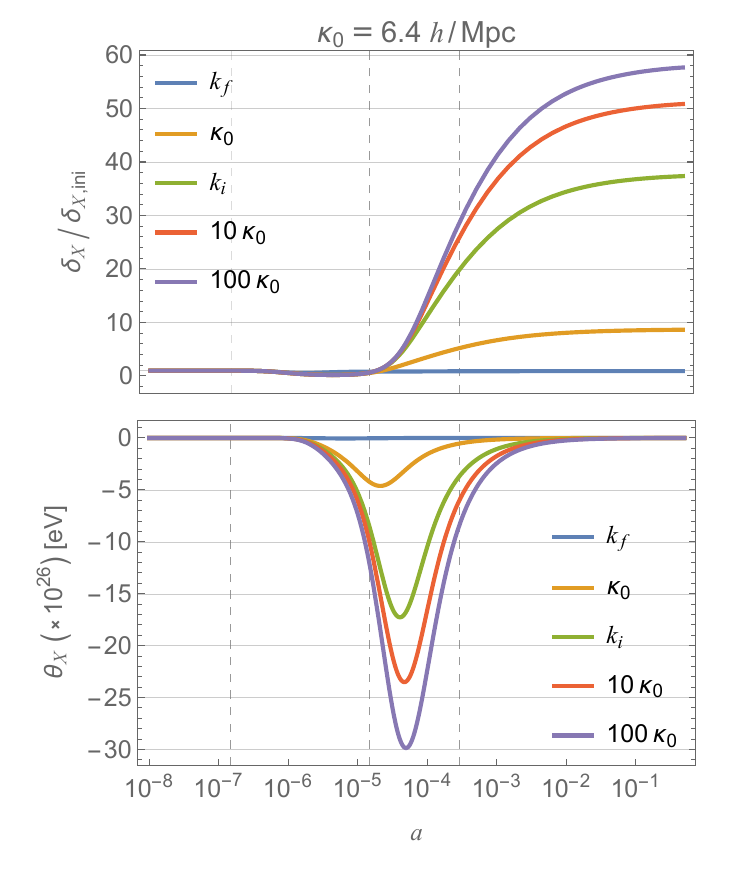}
    \end{subfigure}
    \caption{Diagnostic evolution of the $(\d_X,\theta_X)$ system obtained by setting $\dot h=0$. While all unstable modes exhibit an initial dip, the subsequent growth becomes significantly more pronounced as $\kappa_0$ is increased. The vertical dashed lines denote $a_i,~a_f$, and $a_\rm{eq}$.}
    \label{fig:pert}
\end{figure*}
The GDM fluid perturbation evolution is governed by the effective sound speed squared, $c_s^2\equiv\frac{\d p_X}{\d\rho_X}\big\rvert_\rm{rf}$, where `rf' denotes the GDM rest frame. We adopt the following ansatz:
\begin{equation}\label{cs2}
    c_s^2(a, k) = c_a^2(a)\times\frac{1}{1+(k/a\kappa)^2},
\end{equation}
which ensures that $c_s^2$ vanishes when $c_a^2$ vanishes\footnote{In general, $c_s^2$ and $c_a^2$ are independent quantities and their relation depends on the underlying microphysics. Here, in the absence of a specific microphysical model, we chose the above form as a simple phenomenological parametrization.} (which in turn approximately coincides with the vanishing of $w_X$).  During the transient epoch, $c_a^2$ (and hence $c_s^2$) temporarily becomes negative, which induces an instability that enhances the growth of perturbations for modes which became sub-horizon before or during the instability (see Fig.~\ref{fig:bg}). The $k$-dependent Lorentzian factor acts as an ultraviolet regulator suppressing the effective sound speed for $k/a\gg \kappa$. In this phenomenological framework, $\kappa^{-1}$ represents a physical crossover scale separating the regime in which the fluid exhibits its full adiabatic pressure response from that in which the effective pressure perturbation is progressively suppressed. Additionally, this form of $c_s^2$ yields $c_s^2-c_a^2\propto k^2$ for $k\ll a\kappa$, which occurs, for example, in the case of GDM arising from two interacting adiabatic fluids~\cite{Kopp_2016}. The impact of the instability is governed by the combination $k^2c_s^2$, which scales as $c_a^2k^2$ if $k/a\ll\kappa$ (the adiabatic limit) whereas if $k/a\gg\kappa$, then $c_s^2k^2\to c_a^2a^2\kappa^2$. Consequently the density perturbation grows with $k$ at large scales but saturates on sufficiently small scales. Since the instability essentially terminates at $a_f$, it is convenient to parametrize $\kappa$ through $\kappa=\kappa_0/a_f$. So $\kappa_0$ identifies the comoving scale (evaluated at $a_f$) at which the transition between the adiabatic and the `regulated' regimes occurs.

\begin{figure*}[t]
    \centering
    \begin{subfigure}{0.48\textwidth}
        \centering
        \includegraphics[width=\textwidth]{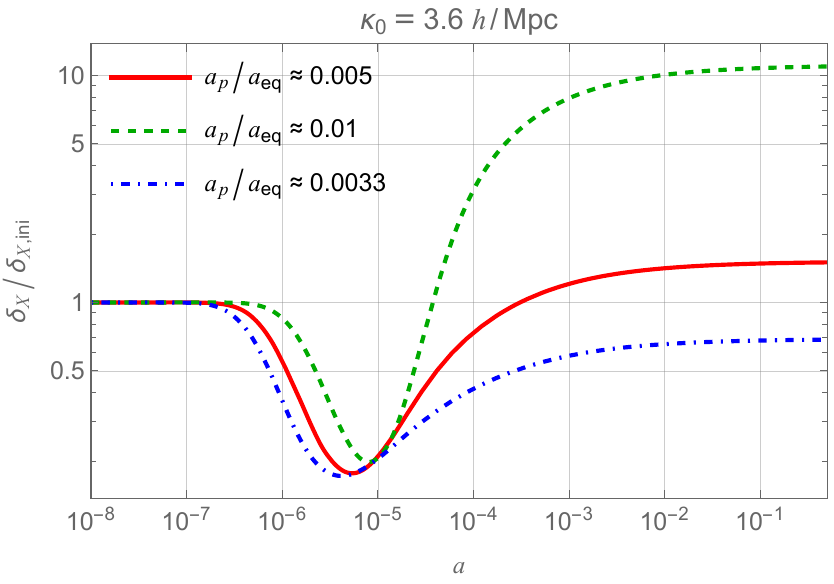}
        \caption{Varying $a_p$ with fixed $\sigma_w$.}
    \end{subfigure}
    \hfill 
    \begin{subfigure}{0.48\textwidth}
        \centering
        \includegraphics[width=\textwidth]{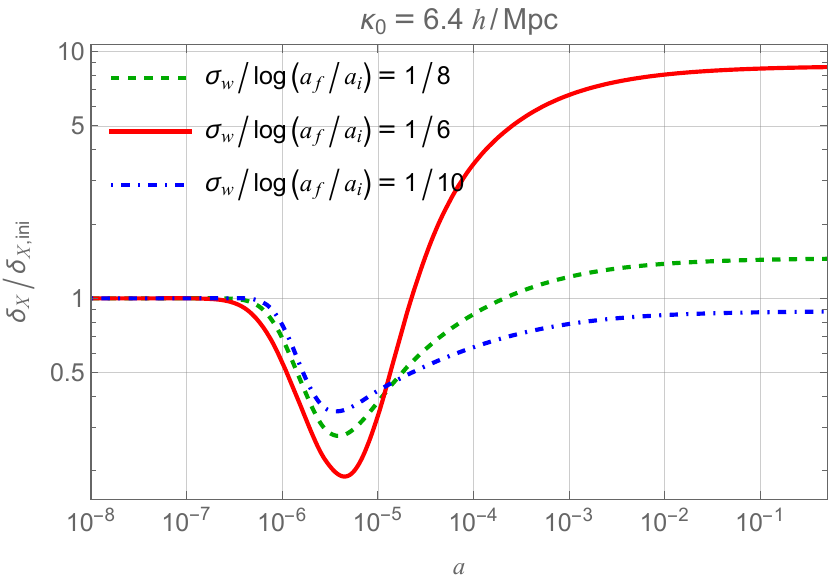}
        \caption{Varying $\sigma_w$ with fixed $a_p$.}
    \end{subfigure}
    \caption{Diagnostic plots for the evolution of the density contrast $\d_X(\kappa_0,a)$ (with $\dot h=0$) while $a_p$ is varied keeping $\sigma_w$ fixed and vice versa. The red solid curves denote the $\{a_p,\sigma_w\}$ combination used in this work.}
    \label{fig:ap_sigmaw}
\end{figure*}

With the expressions for $w_X(a)$, $c_a^2(a)$, and $c_s^2(a,k)$ at hand, we now consider the perturbation equations. We shall work in the synchronous gauge\footnote{The synchronous gauge is characterized by a perturbed line element of the form~\cite{Ma_1995}: 
\begin{align*}
    ds^2=a^2(\tau)\l[-d\tau^2+(\d_{ij}+h_{ij})dx^idx^j\r].
\end{align*}} and in a frame comoving with the standard CDM component in the Universe. The evolution of the density contrast $\d_X=\d\rho_X/\bar\rho_X$ and the velocity divergence $\theta_X$ are governed by the continuity and Euler equations~\cite{PhysRevD.88.127301,PhysRevD.94.023510}:
\begin{align}
    &\dot\d_X+(1+w_X)\l(\theta_X+\frac{\dot h}{2}\r)\nn\\
    &~~~~~~~~~~~~~~~~~~~~~+3\mathcal H\l(\frac{\d p_X}{\d\rho_X}-w_X\r)\d_X=0,\\
    &\dot\theta_X+\mathcal H(1-3w_X)\theta_X+\frac{\dot w_X}{1+w_X}\theta_X\nn\\
    &~~~~~~~~~~~~~~~~~~-\frac{\d p_X/\d\rho_X}{1+w_X}k^2\d_X+k^2\sigma_X=0,
\end{align}
where $h\equiv\d^{ij}h_{ij}$ is the trace of the metric perturbation and
\begin{align}\label{dpX}
    \d p_X&=c_s^2~\d\rho_X-\dot{\bar\rho}_X(c_s^2-c_a^2)\frac{\theta_X}{k^2}\nn\\
    &=c_s^2~\d\rho_X+3\mathcal H(1+w_X)\bar\rho_X(c_s^2-c_a^2)\frac{\theta_X}{k^2}.
\end{align}
We neglect anisotropic stress in the GDM sector and set $\sigma_X=0$, which is equivalent to setting $c_\rm{vis}^2$, which governs the evolution of $\sigma_X$, to zero~\cite{PhysRevD.94.023510}. The perturbation equations therefore take the following form:
\begin{align}
    \dot\d_X&=-(1+w_X)\l(\theta_X+\frac{\dot h}{2}\r)-3\mathcal H(c_s^2-w_X)\d_X\nn\\
    &~~~~~~~~~~~~~~~~~~~~~~-9(1+w_X)(c_s^2-c_a^2)\frac{\mathcal H^2}{k^2}\theta_X,\\
    \dot\theta_X&=-(1-3c_s^2)\mathcal H\theta_X+\frac{c_s^2k^2}{1+w_X}\d_X.
\end{align}
In order to close these equations we need to use the linearized Einstein equation for $h$:
\begin{align}
    \ddot h+\mathcal{H}\dot h&=-\frac{a^2}{M_P^2}\bigg[2\bar\rho_r\d_r+\bar\rho_b\d_b+\bar\rho_{c}\d_{c}+(1+3c_s^2)\bar\rho_X\d_X\nn\\
    &~~~~~~~~~~~~+9\mathcal H(1+w_X)(c_s^2-c_a^2)\frac{\bar\rho_X\theta_X}{k^2}\bigg],
\end{align}
where $M_P\equiv1/\sqrt{8\pi G}$ denotes the reduced Planck mass and the subscripts `$r$', `$b$', and `$c$' stand for radiation, baryons and CDM, respectively. 

Even though the exotic DM component becomes indistinguishable from CDM at the background level once the instability shuts down ($w_X\simeq c_s^2\simeq0$), the same is not true at the level of linear perturbations. For $a\gtrsim a_f$, the perturbation equations simplify to
\begin{align}
    \dot\d_X=-\theta_X-\frac{\dot h}{2},\quad \dot\theta_X=-\mathcal H\theta_X,
\end{align}
with $\theta_X$ not identically zero, unlike that for CDM. Consequently, the perturbations retain a memory of the earlier instability through the residual velocity field, leading to an evolution that differs from that of CDM even after $w_X$ and $c_s^2$ have become negligible. Before studying the full solution, it is useful to isolate the effect of the transient instability from the gravitational forcing associated with metric perturbations. To this end, we temporarily set $\dot h=0$ and evolve the coupled $(\d_X,\theta_X)$ system starting with the usual CDM adiabatic initial conditions imposed at $a\ll a_i$. The resulting evolution of $\delta_X$ and $\theta_X$  is shown in Fig.~\ref{fig:pert}. In Fig.~\ref{fig:ap_sigmaw} we plot $\d_X(\kappa_0,a)$ for different values of $a_p$ and $\sigma_w$. As $a_p$ shifts towards the epoch of matter-radiation equality, a larger fraction of the relevant modes are already sub-horizon during the period when $|c_s^2|$ is maximal. Consequently, the resulting enhancement of $\d_X(k,a)$ becomes more pronounced. On the other hand, if the instability window shrinks while keeping $a_p$ fixed, the recovery (from the initial dip) is weak/incomplete, as expected.

During the transient epoch, the direct pressure contribution $-3\mathcal H(c_s^2-w_X)\d_X$ initially dominates the evolution of $\d_X$, driving a suppression of the density contrast and producing the dip seen in the upper panels of Fig.~\ref{fig:pert}. Simultaneously, the negative sound speed squared generates a non-zero velocity divergence through the Euler equation, causing $\theta_X$ to become increasingly negative. As $|\theta_X|$ grows, the continuity equation contribution $-(1+w_X)\theta_X$ becomes increasingly important and eventually overcomes the pressure term. The minimum of $\d_X$ occurs when these competing contributions approximately balance each other. Thereafter the velocity term dominates, leading to a recovery and subsequent growth of the density contrast. The efficiency of this recovery depends on the magnitude of the velocity divergence generated during the transient, which increases with $k$ unless $k\gg\kappa_0$. Consequently, modes with sufficiently large $k$ (which determines how much time a mode spends in the instability window) and/or larger values of $\kappa_0$ experience a net enhancement of $\d_X$, whereas for smaller $k$ or smaller $\kappa_0$ the recovery remains incomplete and the final density contrast stays below its initial value. From the two plots we also observe that the relative growth between $k=10\kappa_0$ and $100\kappa_0$ is significantly smaller than that between $k=\kappa_0$ and $10\kappa_0$ reflecting the near-saturation behavior of $c_s^2k^2$ for $k\gg\kappa_0$. 

Having built the intuition for the instability-driven growth mechanism using the diagnostic $\dot h=0$ computation, we now turn to the full cosmological evolution in the next section.

\section{Cosmological Observables: from $P(k,z)$ to UV LF}\label{sec:Pk_UVLF}
\begin{figure}
    \centering
    \includegraphics[width=1.05\linewidth]{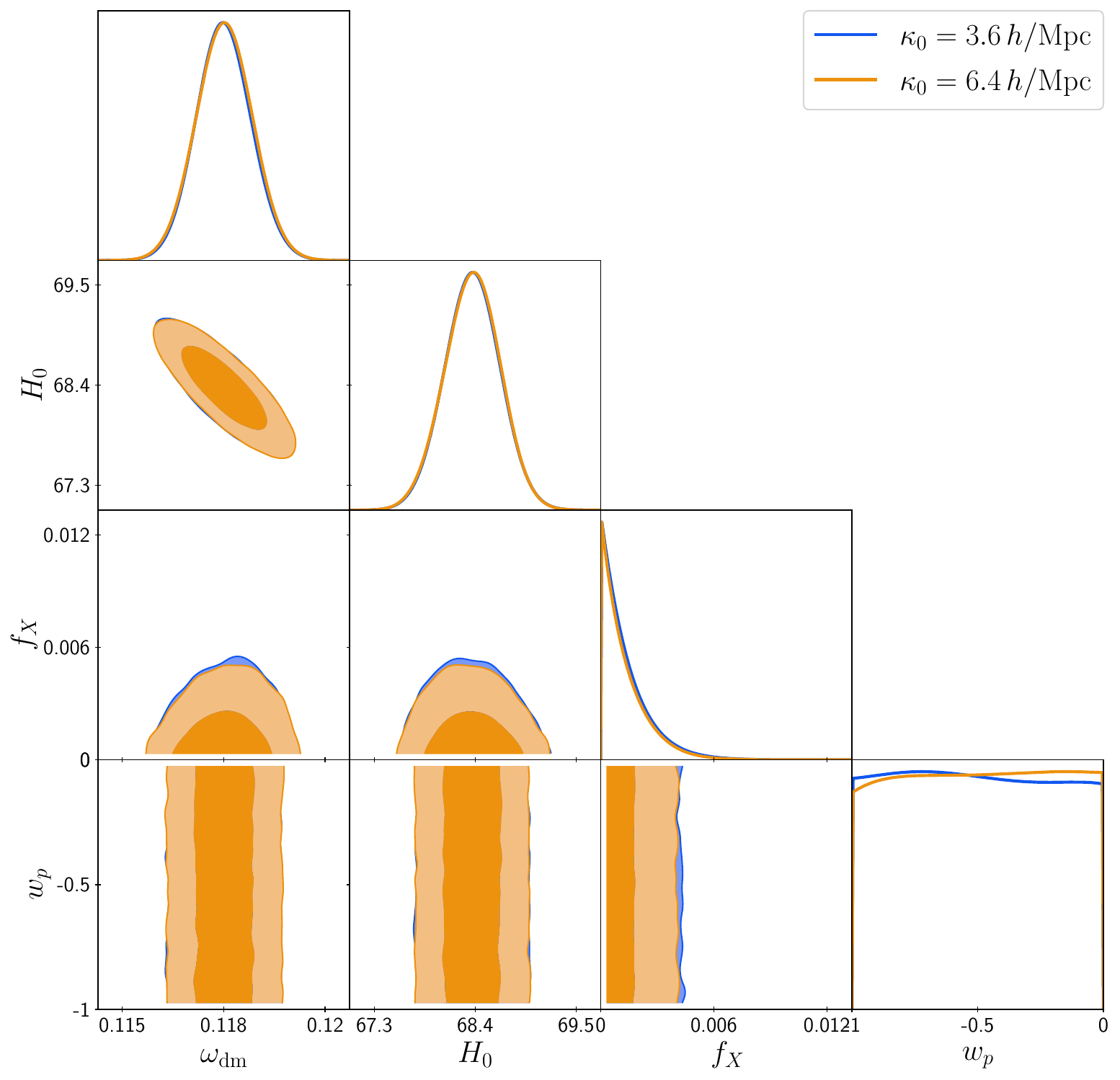}
    \caption{1D and 2D marginalised posterior distributions obtained for $\kappa_0=3.6\,h/\mathrm{Mpc}$ and $\kappa_0=6.4\,h/\mathrm{Mpc}$ from MCMC analysis using combined dataset: Planck + DESI DR2 + Pantheon+. The contours correspond to the $68\%$ (dark shaded) and $95\%$ (light shaded) confidence regions.}
    \label{fig:gdm_mcmc}
\end{figure}
\begin{figure*}[htbp]
    \centering
    \includegraphics[width=0.9\linewidth]{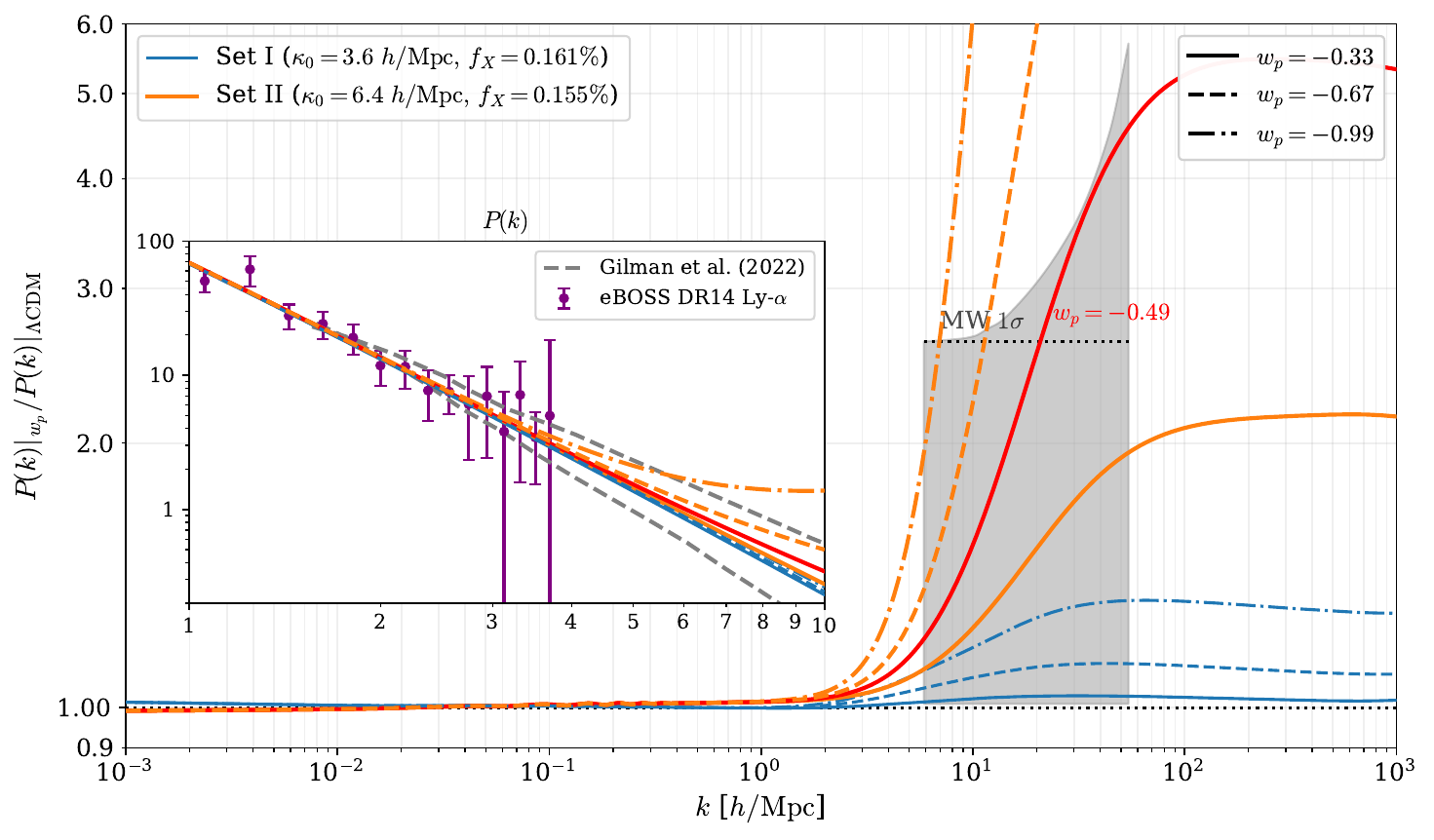}
    \caption{Main: Ratio of the modified and the $\La$CDM matter power spectra for different values of $w_p$, at $z=0$, with $f_X$ fixed to its marginalized $95\%$ C.L. upper limit for each benchmark set. The gray shaded region corresponds to the $1\sigma$ allowed region from an analysis of Milky Way dwarf galaxies, and the dotted line is its geometric mean~\cite{esteban2024milkywaysatellitevelocities}. Inset: The matter power spectra in the $k$-range $1-10~h/\rm{Mpc}$ along with SDSS $\rm{Ly}\a$ constraints~\cite{10.1093/mnras/stz2310} and $1\sigma$ constraints from strong lensing shown by gray dashed lines~\cite{10.1093/mnras/stac670}.}
    \label{fig:kappa}
\end{figure*}
\subsection{The Matter Power Spectrum}
In order to solve the full set of equations and obtain the total linear matter power spectrum, we modified the fluid module in the publicly available package \texttt{CLASS}~\cite{Diego_Blas_2011}. For a fixed $\kappa_0$, the impact of $X$ on the primary CMB is governed mainly by $f_X$ (which essentially determines how much of standard CDM is replaced by GDM) and $w_p$. To determine the allowed parameter space of the model, we perform a Markov Chain Monte Carlo (MCMC) analysis using a combination of CMB, BAO, and Type Ia supernova measurements. The datasets included in our analysis are summarized below:

\begin{itemize}
\item[$\blacktriangleright$] \textbf{Planck \hspace{0.5mm}2018:} We used the Planck 2018 CMB likelihood, including the high-multipole temperature and polarization spectra (TT, TE, and EE), together with the low-multipole temperature and $E$-mode polarization measurements \cite{2020}. The corresponding Planck likelihoods were evaluated using the official \texttt{clik} likelihood framework.
\vspace{2mm}
\item[$\blacktriangleright$]\textbf{DESI DR2 BAO:} We included baryon acoustic oscillation measurements from the second DESI data release \cite{2025desidr2}. The sample combines several large-scale-structure tracers, including galaxies, quasars, and the Lyman-$\alpha$ forest, and provides isotropic and anisotropic BAO information over the redshift interval $0.295 \leq z \leq 2.330$, distributed across nine redshift bins as listed in Table IV of Ref.~\cite{2025desidr2}.
\vspace{2mm}
\item[$\blacktriangleright$] \textbf{Pantheon+ :} We incorporated the Pantheon+ compilation of Type Ia supernovae, which probes the late-time expansion history over the redshift range
$0.001 \lesssim z \lesssim 2.26$ \cite{Brout_2022}.
\end{itemize}

We confronted our model with the above data sets, for two benchmark values of $\kappa_0$, namely $\kappa_0=3.6~h/\rm{Mpc}$ and $\kappa_0=6.4~h/\rm{Mpc}$. We performed the MCMC sampling with the publicly available \texttt{MontePython-v3} package \cite{brinckmann2018montepython}, interfaced with our modified \texttt{CLASS} implementation, using the Metropolis--Hastings algorithm with Cholesky-based parameter decomposition to improve sampling efficiency \cite{Lewis_2000}. Chain convergence was required to satisfy the Gelman--Rubin criterion $R-1<0.01$ for all parameters \cite{gelman1992inference}, while the minimum $\chi^2$ was determined using the SLSQP optimizer available in \texttt{MontePython}.

Our baseline cosmological model includes the six standard $\La$CDM parameters: $\{\w_b, \w_\rm{dm}, H_0, A_s, n_s, \tau_\rm{reio}\}$ (with the $\w_c$ parameter replaced by $\w_\rm{dm}$, the total DM density), along with the GDM parameters $w_p$ and $f_X$. The physical CDM and exotic DM densities are given by $\w_c=(1-f_X)\w_\rm{dm}$ and $\w_X=f_X\w_\rm{dm}$, respectively. In the limit $f_X\to0$ and $w_p\to0$ (and hence $c_s^2\propto c_a^2\to0$), we recover $\La$CDM cosmology. We adopted flat priors for the standard parameters, except for $\tau_\rm{reio}$, for which we imposed a lower bound of $0.004$. The additional model parameters are assigned uniform priors $f_X\in[0.00,0.99]$ and $w_p\in[-0.99,0.00]$. The results of the MCMC analysis are presented in Table~\ref{tab:mcmc_res} and shown in Fig.~\ref{fig:gdm_mcmc}. We find that the standard cosmological parameters remain close to their $\La$CDM values, while the data places stringent upper limits on $f_X$. In contrast, $w_p$ remains effectively unconstrained. This is because once $f_X$ is constrained to be small, substantially different values of $w_p$ can produce nearly indistinguishable CMB spectra. Additionally, the corresponding 2D marginalized posteriors show no appreciable correlation of $w_p$ with either the standard cosmological parameters or $f_X$. The minimum $\chi^2$ values for Set I and Set II, listed in Table~\ref{tab:mcmc_res}, are only marginally lower than that for $\La$CDM, indicating no significant preference for the extended model from these data sets.
\begin{table*}[htbp]
\centering
\setlength{\tabcolsep}{12pt}
\begin{tabular}{c|c|c|c}
\hline\hline
Parameter & $\Lambda$CDM & \multicolumn{1}{c|}{Set I} & Set II \\
& & $(\kappa_0=3.6\,h/\mathrm{Mpc})$ &
$(\kappa_0=6.4\,h/\mathrm{Mpc})$ \\
\cline{1-4}

$100\,\omega_b$
& $2.25(2.26)\pm0.0131$
& $2.25(2.24)^{+0.0131}_{-0.0130}$
& $2.25(2.25)^{+0.013}_{-0.0127}$ \\

$\omega_\rm{ dm}$
& $0.118(0.118)^{+0.000665}_{-0.000688}$
& $0.118(0.118)^{+0.000663}_{-0.000681}$
& $0.118(0.118)^{+0.000675}_{-0.000687}$ \\

$H_0$
& $68.4(68.5)\pm0.302$
& $68.3(68.3)^{+0.300}_{-0.301}$
& $68.3(68.2)^{+0.301}_{-0.306}$ \\

$\ln(10^{10}A_s)$
& $3.05(3.04)^{+0.0160}_{-0.0177}$
& $3.05(3.05)^{+0.0161}_{-0.0179}$
& $3.05(3.05)^{+0.016}_{-0.018}$ \\

$n_s$
& $0.971(0.972)\pm0.00342$
& $0.971(0.968)^{+0.00347}_{-0.00344}$
& $0.971(0.973)^{+0.00345}_{-0.00342}$ \\

$\tau_\rm{reio}$
& $0.0575(0.058)^{+0.00764}_{-0.00845}$
& $0.058(0.059)^{+0.00762}_{-0.00870}$
& $0.058(0.056)^{+0.00758}_{-0.00865}$ \\

\hline

$f_X$
& $\rm{N/A}$
& $<0.00393$ 
& $<0.00376$ 
\\

$w_p$
& $\rm{N/A}$
& $-$ 
& $-$ 
\\
\hline
$\chi_\rm{min}^2$ & $4200.94$ & $4200.497$ & $4200.057$\\
\hline\hline
\end{tabular}

\caption{Comparison of cosmological parameter constraints from
Planck+DESI DR2+Pantheon+ data. The quoted values are $\text{mean(best-fit)}\pm1\sigma$ for the $\Lambda$CDM parameters, while for $f_X$ we report the $95\%$ C.L. upper bounds. A dash in the Set I and Set II columns indicates that $w_p$ is unconstrained by the data.}
\label{tab:mcmc_res}
\end{table*}

In Fig.~\ref{fig:kappa}, we show the ratio of the modified matter power spectrum for different $w_p$ values and the $\La$CDM matter power spectrum. The gray shaded region corresponds to constraints from Milky Way (MW) dwarf galaxies~\cite{esteban2024milkywaysatellitevelocities}. The shape of the total linear matter power spectrum is determined by the interplay between the non-zero velocity divergence-driven growth and the subsequent gravitational backreaction. Once $c_s^2k^2\simeq c_a^2a^2\kappa_0^2$, the impact of metric sourcing takes over, leading to a turnover/saturation depending on the values of $\kappa_0$ and $w_p$. 
For $\kappa_0=3.6~h/\rm{Mpc}$, all values of $w_p>-1$ satisfy the MW $1\sigma$ bounds. Similarly, for $\kappa_0=6.4~h/\rm{Mpc}$, all $w_p\gtrsim-0.49$ (shown in red) are permitted by the MW $1\sigma$ constraint. As illustrated by graphical overlays (see inset), both of these restricted $w_p$ parameter spaces also exhibit broad qualitative consistency with existing Lyman-$\alpha$ and strong lensing limits~\cite{10.1093/mnras/stac670,10.1093/mnras/stz2310}. The figure illustrates the maximal matter power spectrum deviation allowed by the cosmological analysis, obtained by fixing $f_X$ to its marginalized $95\%$ C.L. upper bound, $f_X^{95}=0.00393$ for Set I and $f_X^{95}=0.00376$ for Set II. In the subsequent analysis, we shall instead adopt the corresponding $68\%$ C.L. upper limit values, $f_X^{68}=0.00161$ and $0.00155$ for Sets I and II, respectively, unless otherwise specified. For the latter case, we found that the Set I matter power spectrum remains essentially indistinguishable from $\La$CDM over the full range $w_p\in[-0.99,0]$, while for Set II, the MW constraint allows $w_p\gtrsim-0.65$ only. 
The two benchmark choices of $\kappa_0$, viz. $\kappa_0=3.6~h/\rm{Mpc}$ and $\kappa_0=6.4~h/\rm{Mpc}$, illustrate a case with only marginal improvement over $\La$CDM (even for the extreme limiting value $w_p\approx-0.99$), and one with a substantial enhancement of early structure formation and a correspondingly lower inferred star-formation efficiency, as we shall see below. We explore the dependence of our results on the choice of $\kappa_0$ beyond the two representative benchmark values in Appendix~\ref{sec:kappa0}. 

\subsection{The Ultraviolet Luminosity Function}
The ultraviolet luminosity function (UV LF) is related to the underlying matter power spectrum indirectly through the halo mass function. The halo mass function (HMF) is defined as the (comoving) number density ($n$) of DM halos per unit mass:
\begin{align}\label{hmf}
    \frac{dn}{d\ln{M_h}}=f(\sigma)\frac{\bar\rho_{m,0}}{M_h}\bigg|\frac{d\ln{\sigma}}{d\ln{M_h}}\bigg|,
\end{align}
where $\sigma(R)$ and $\bar\rho_{m,0}$ are the mass variance of smoothed linear matter density field in a sphere of radius $R$ and the mean matter density of the Universe, respectively. The sphere of radius $R$ encloses a mass $M_h=4\pi\bar\rho_{m,0}R^3/3$ within it. The mass variance depends on the linear matter power spectrum $P(k,z)$ through
\begin{align}
    \sigma^2(R,z)=\frac{1}{2\pi^2}\int_0^\infty k^2P(k,z)W^2(kR)~dk,
\end{align}
where $k$ is the wavenumber and $W(kR)$ is a filter function in Fourier space, which we take to be of the top-hat form:
\begin{align}
    W(kR)=3\frac{\sin{(kR)}-kR\cos{(kR)}}{(kR)^3}.
\end{align}
We use the Sheth-Tormen fitting function $f(\sigma)$, obtained using the Press-Schechter formalism~\cite{1974ApJ...187..425P} and including the corrections for ellipsoidal collapse~\cite{ST}:
\begin{align}\label{st}
    f(\sigma)=A\sqrt{\frac{2a}{\pi}}\l[1+\l(\frac{\sigma^2}{a\d_c^2}\r)^p\r]\frac{\d_c}{\sigma}\exp{\l(-\frac{a\d_c^2}{2\sigma^2}\r)},
\end{align}
where $\d_c\approx1.686$ is the critical overdensity for gravitational collapse, $A=0.3222,~a=0.707,$ and $p=0.3$.

To connect the halo abundance to observable galaxy populations, one needs to employ a galaxy-halo connection prescription that relates the halo mass $M_h$ to the star-formation rate (SFR) of the galaxy it hosts. The resulting SFR is then converted into a UV luminosity, allowing the UV LF to be constructed from the HMF. The SFR is expressed as~\cite{harikane2023purespectroscopicconstraintsuv}
\begin{align}
    \rm{SFR}=f_\rm{SF}\times f_b\times\frac{dM_h}{dt}(M_h,z),
\end{align}
where $f_\rm{SF}$ is the star-formation efficiency, $f_b=\O_b/\O_m=0.157$ is the cosmic baryon fraction and $dM_h/dt$ is the matter accretion rate. Following Ref.~\cite{Yung_2025}, we use the following expression for the accretion rate\footnote{\label{footnote:yung}In the absence of dedicated simulations for the present model, we assume that the halo accretion rate and galaxy-halo connection are adequately described by the standard $\La$CDM-calibrated prescription. The primary effect of the GDM model is therefore assumed to arise from the modified HMF.},
\begin{align}
    \frac{dM_h/dt}{M_\odot~\rm{yr}^{-1}}&=\b(z)\l[\frac{M_h}{10^{12}M_\odot}\frac{H(z)}{H_0}\r]^{\a(z)},\\
    \a(z)&=0.948+0.694a-0.565a^2,\nn\\
    \log{\b(z)}&=2.673-2.075a+0.891a^2,\nn
\end{align}
with $a=1/(1+z)$. We use the following conversion equation between SFR and UV luminosity $L_\rm{UV}=\rm{SFR}/\kappa_\rm{UV}$~\cite{Sabti_2022} with $\kappa_\rm{UV}=0.72\times10^{-28}~\rm{M_\odot~yr^{-1}~erg^{-1}~s~Hz}$, assuming Chabrier IMF~\cite{Chabrier_2003,Madau_2014}. The absolute magnitude and the luminosity are related by
\begin{align}
    M_\rm{UV}=-2.5\log{\l(\frac{L_\rm{UV}}{\rm{erg~s^{-1}~Hz^{-1}}}\r)}+51.63.
\end{align}
Owing to the stochastic nature of galaxy formation, we model the $M_\rm{UV}-M_h$ relation using a Gaussian kernel centered about the median $\hat M_\rm{UV}$ and with a width $\sigma_\rm{UV}$~\cite{das2025darksecretsbaryonsilluminating}:
\begin{align}
    \Pi(M_\rm{UV}|M_h)=\frac{1}{\sqrt{2\pi}\sigma_\rm{UV}}\exp{\l(-\frac{(M_\rm{UV}-\hat M_\rm{UV})^2}{2\sigma^2_\rm{UV}}\r)},
\end{align}
which represents the probability that a halo of mass $M_h$ hosts a galaxy of absolute magnitude $M_\rm{UV}$. Following Ref.~\cite{harikane2023purespectroscopicconstraintsuv}, we assume a $0.2$ dex scatter in the halo mass and set $\sigma_\rm{UV}=2.5\times\a(z)\times0.2$. The UV LF is then computed as
\begin{align}
    \Phi(M_\rm{UV})=\int d\ln{M_h}~\l(\frac{dn}{d\ln{M_h}}\r)\Pi(M_\rm{UV}|M_h).
\end{align}
We discuss our results in the next section.

\begin{figure}[htbp]
    \centering
    \begin{subfigure}{0.48\textwidth}
        \centering
        \includegraphics[width=\textwidth]{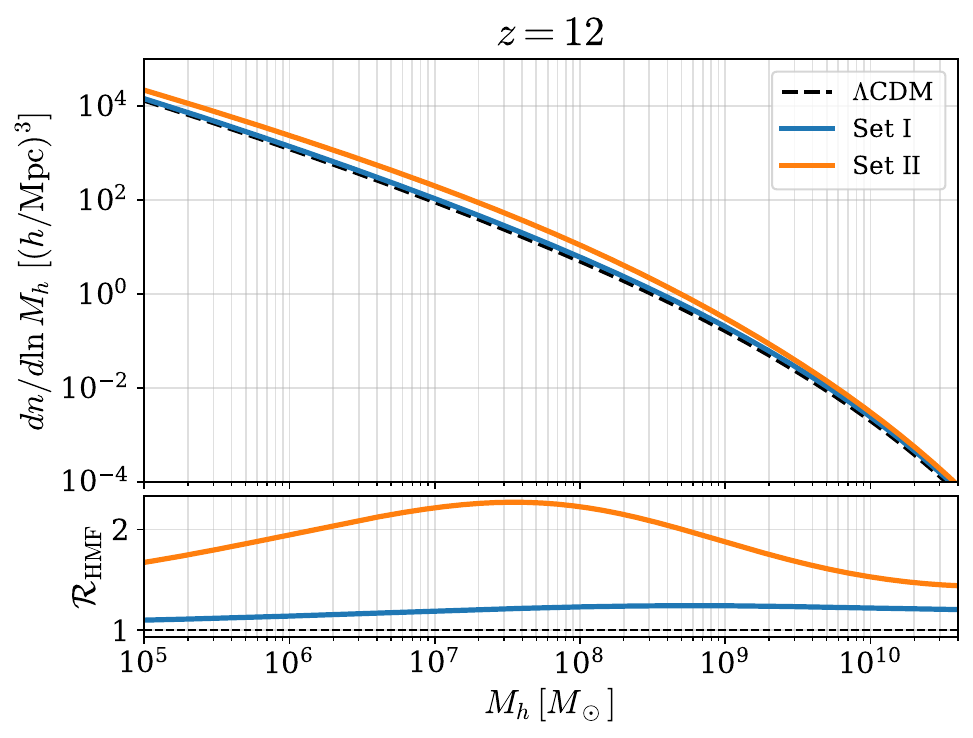}
    \end{subfigure}
    \hfill 
    \begin{subfigure}{0.48\textwidth}
        \centering
        \includegraphics[width=\textwidth]{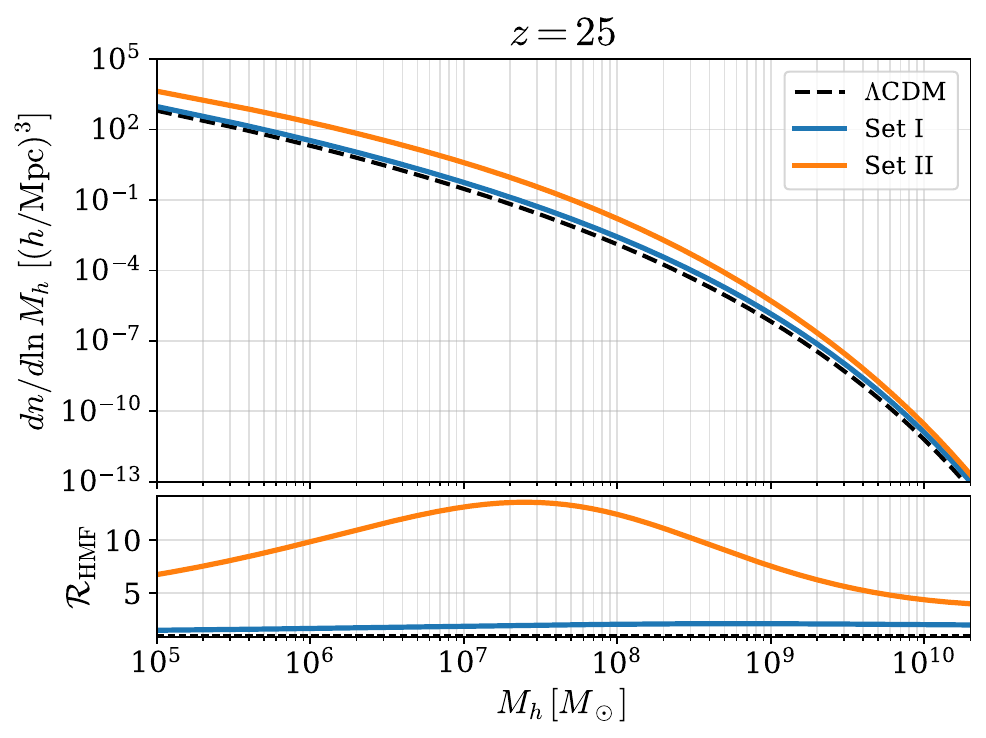}
    \end{subfigure}
    \caption{Halo mass function (top panels) for our model and its enhancement relative to $\La$CDM (bottom panels) at $z=12$ and $z=25$. The enhancement in Set II extends into the halo mass range expected to host the brightest high-redshift galaxies, suggesting that it should have a greater impact on the predicted abundance of luminous JWST galaxies.}
    \label{fig:HMF}
\end{figure}

\section{Results} \label{sec:results}

\begin{figure*}[p]
\centering

\begin{minipage}{\textwidth}
\centering
\setlength{\tabcolsep}{10pt}

\begin{tabular}{c|c|c|c|c|c|c}
\hline\hline

\multirow{3}{*}{$z$}
& \multirow{3}{*}{$\Lambda$CDM}
& \multicolumn{2}{c|}{Set I: $\kappa_0=3.6\,h/\mathrm{Mpc}$}
& \multicolumn{3}{c}{Set II: $\kappa_0=6.4\,h/\mathrm{Mpc}$}
\\
\cline{3-4}
\cline{5-7}

&
& $f_X^{68}=0.00161$
& $f_X^{95}=0.00393$
& \multicolumn{2}{c|}{$f_X^{68}=0.00155$}
& $f_X^{95}=0.00376$
\\
\cline{3-7}

&
& $w_p=-0.99$
& $w_p=-0.99$
& $w_p=-0.33$
& $w_p=-0.45$
& $w_p=-0.33$
\\

\hline

12
& $0.056_{-0.021}^{+0.030}$
& $0.052_{-0.020}^{+0.028}$
& $0.049_{-0.018}^{+0.025}$
& $0.050_{-0.019}^{+0.026}$
& $0.049_{-0.018}^{+0.026}$
& $0.048_{-0.018}^{+0.025}$
\\

14
& $0.081_{-0.024}^{+0.024}$
& $0.075_{-0.022}^{+0.021}$
& $0.069_{-0.020}^{+0.019}$
& $0.072_{-0.021}^{+0.020}$
& $0.069_{-0.020}^{+0.019}$
& $0.067_{-0.019}^{+0.018}$
\\

\hline

17
& $0.045_{-0.012}^{+0.009}$
& $0.039_{-0.011}^{+0.008}$
& $0.033_{-0.009}^{+0.006}$
& $0.036_{-0.010}^{+0.007}$
& $0.032_{-0.008}^{+0.006}$
& $0.030_{-0.008}^{+0.006}$
\\

25
& $0.187_{-0.072}^{+0.043}$
& $0.147_{-0.056}^{+0.033}$
& $0.109_{-0.042}^{+0.024}$
& $0.121_{-0.046}^{+0.026}$
& $0.091_{-0.034}^{+0.019}$
& $0.083_{-0.031}^{+0.017}$
\\

\hline\hline
\end{tabular}

\captionof{table}{
best-fit star-formation efficiencies $f_\rm{SF}$ obtained from fitting the UV LFs at different redshifts for $\Lambda$CDM and the two benchmark parameter sets. For Set I, we consider the $68\%$ and $95\%$ C.L. upper limits on $f_X$, with $w_p=-0.99$. For Set II, at the $68\%$ C.L. upper limit on $f_X$ we consider $w_p=-0.33$ and $-0.45$, while at the $95\%$ C.L. upper limit only $w_p=-0.33$ is considered. The quoted uncertainties correspond to the $1\sigma$ confidence intervals derived from $\Delta\chi^2=1$.
}
\label{tab:fsf_bestfit}

\end{minipage}

\vspace{0.35cm}

\captionsetup{type=figure}

\begin{subfigure}[b]{0.48\textwidth}
\centering
\includegraphics[width=\textwidth]{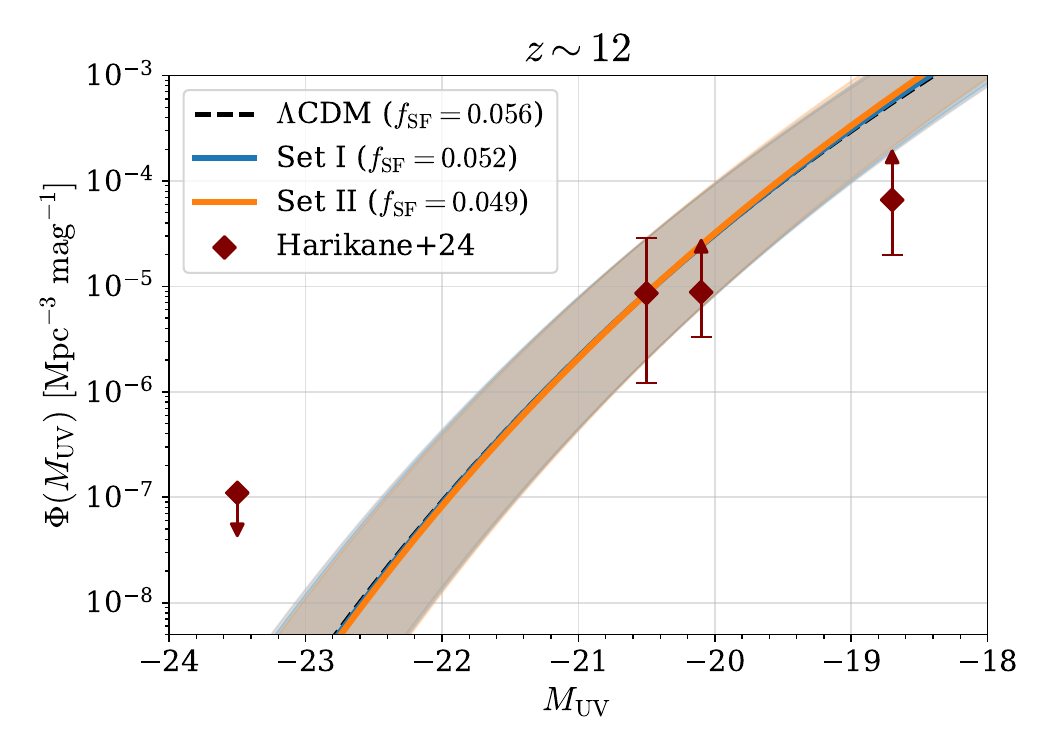}
\end{subfigure}
\hfill
\begin{subfigure}[b]{0.48\textwidth}
\centering
\includegraphics[width=\textwidth]{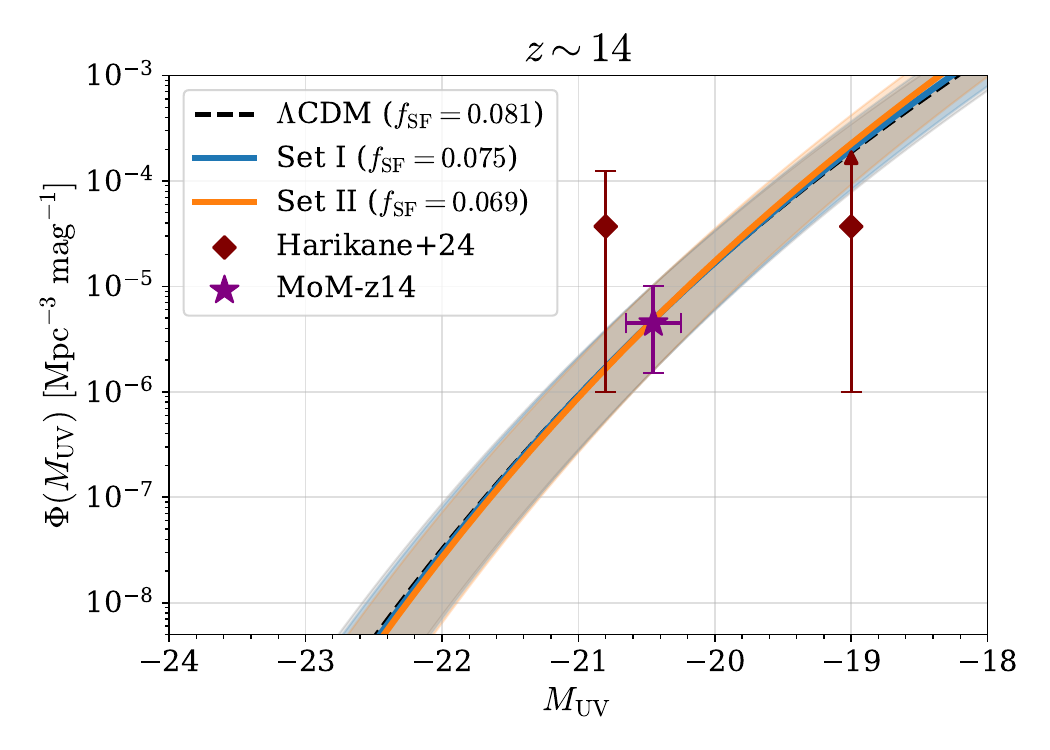}
\end{subfigure}

\vspace{0.1cm}

\begin{subfigure}[b]{0.48\textwidth}
\centering
\includegraphics[width=\textwidth]{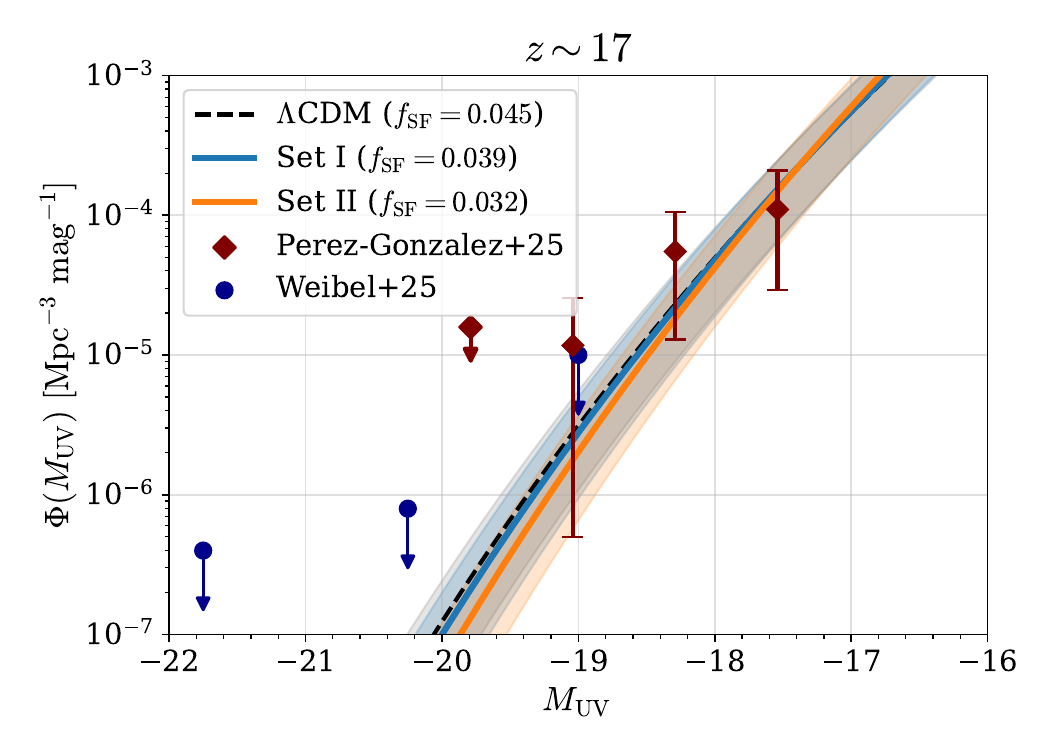}
\end{subfigure}
\hfill
\begin{subfigure}[b]{0.48\textwidth}
\centering
\includegraphics[width=\textwidth]{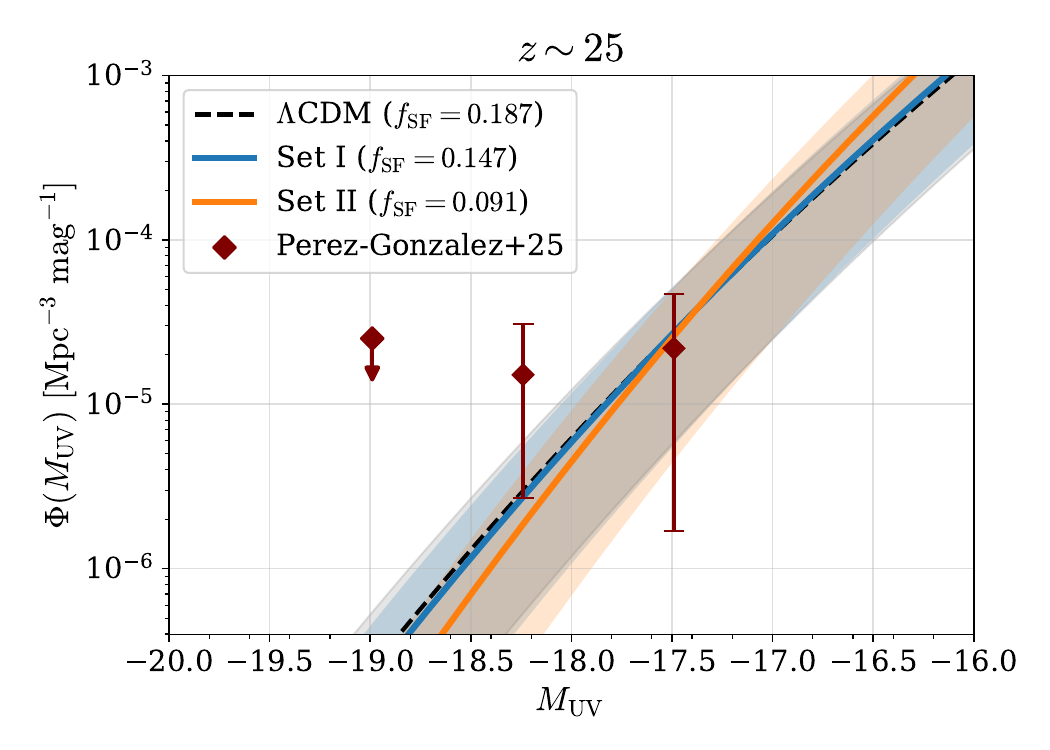}
\end{subfigure}

\caption{Comparison of the predicted UV LFs for our model, with $w_p=-0.99$ for Set I and $w_p=-0.45$ for Set II, using the corresponding $f_X^{68}$ values, and for the $\Lambda$CDM model. For each curve, the star-formation efficiency $f_\rm{SF}$ is chosen to satisfy the observational constraints at the corresponding redshift. The data points correspond to spectroscopically confirmed galaxies at $z\sim12,14$~\cite{harikane2024jwstalmakeckspectroscopic, naidu2026cosmicmiracleremarkablyluminous} and photometric candidates at $z\sim17,25$~\cite{Weibel_2026, perezgonzalez2025risegalacticempireluminosity}. The shaded regions denote the $1\sigma$ uncertainties in $f_\rm{SF}$.}
\label{fig:main_res}
\end{figure*}

The dominant impact of the exotic species is on the halo abundance. The conversion from halo abundance to UV LF requires additional astrophysical assumptions, which we model using the standard galaxy-halo connection described in Sec.~\ref{sec:Pk_UVLF}. In Figure~\ref{fig:HMF}, we show the HMFs at $z=12$ and $z=25$ along with the corresponding ratios $\mathcal R_\rm{HMF}\equiv\frac{\text{HMF}|_{w_p}}{\text{HMF}|_{\La\rm{CDM}}}$. Both benchmark models (Set I with $w_p=-0.99$ and Set II with $w_p=-0.45$) predict only a modest enhancement in the halo abundance at $z=12$, while the enhancement becomes substantially larger by $z=25$. The ratio plots further reveal that the HMF enhancement is strongly mass dependent. For Set I, the HMF is only mildly enhanced relative to $\La$CDM, with the ratio remaining close to unity over the full mass range shown. Set II, by contrast, exhibits a much stronger enhancement, peaking around $M_h\simeq(2-3)\times10^7M_\odot$, and remains appreciable for nearly five decades in halo mass. We therefore expect Set II to have the larger impact on the abundance of the high-redshift galaxies probed by JWST. This is indeed what we observe, as reported in Table~\ref{tab:fsf_bestfit}. For each redshift and cosmological model, the best-fit star-formation efficiency, $f_\rm{SF}$, is obtained by minimizing an effective $\chi^2$ function over the range $0.001\leq f_\rm{SF}\leq1$ using the bounded Brent minimization algorithm. The effective $\chi^2$ is constructed by comparing the predicted UV LF with the observational constraints (from Harikane et al.~\cite{harikane2024jwstalmakeckspectroscopic}, Naidu et al.~\cite{naidu2026cosmicmiracleremarkablyluminous}, and Perez-Gonzalez et al.~\cite{perezgonzalez2025risegalacticempireluminosity}), with standard Gaussian contributions for detected data points using the appropriate asymmetric uncertainties:
\begin{align}\label{chi2}
    \chi^2_\rm{eff}&=\sum_{i\in\rm{detect}}\frac{[\Phi_{\rm{th},i}-\Phi_i]^2}{\sigma_i^2}\nn\\
    &~+\sum_{i\in\rm{upper}}\Theta(\Phi_{\rm{th},i}-\Phi_i^\rm{lim})\l(\frac{\Phi_{\rm{th},i}-\Phi_i^\rm{lim}}{\Phi_i^\rm{lim}}\r)^2\nn\\
    &~+\sum_{i\in\rm{lower}}\Theta(\Phi_i^\rm{lim}-\Phi_{\rm{th},i})\l(\frac{\Phi_{\rm{th},i}-\Phi_i^\rm{lim}}{\sigma_i^\rm{lim}}\r)^2,
\end{align}
where for detected points $\sigma_i=\sigma_{i,+}$ when $\Phi_{\rm{th},i}>\Phi_i$ and $\sigma_i=\sigma_{i-}$ otherwise. Upper and lower limits are incorporated through one-sided penalty terms, using the Heaviside step function $\Theta$, contributing only when the theoretical prediction $\Phi_{\rm{th},i}\equiv\Phi_\rm{th}(M_{\rm{UV},i})$ violates the corresponding observational bound $\Phi_i^\rm{lim}$. For upper limits, where no uncertainty is reported, the penalty is normalized by the quoted upper limit itself. 
In Fig.~\ref{fig:main_res} we plot the UV LF $\Phi(M_\rm{UV})$ and compare number densities resulting from our model at $z\sim12-25$ with the number densities corresponding to $\La$CDM. Data points (maroon diamonds) represent UV LF measurements of spectroscopically confirmed galaxies at $z\sim12,14$ as derived in Harikane et al.~\cite{harikane2024jwstalmakeckspectroscopic} along with the data point (purple star) from the MoM-z14 galaxy observation~\cite{naidu2026cosmicmiracleremarkablyluminous}, and measurements of photometric galaxy candidates at $z\sim17,25$ as derived in Weibel et al. (dark blue circles)~ \cite{Weibel_2026} and Perez-Gonzalez et al. (maroon diamond)~\cite{perezgonzalez2025risegalacticempireluminosity}. 

\begin{figure}[htbp]
    \centering
    \includegraphics[width=\linewidth]{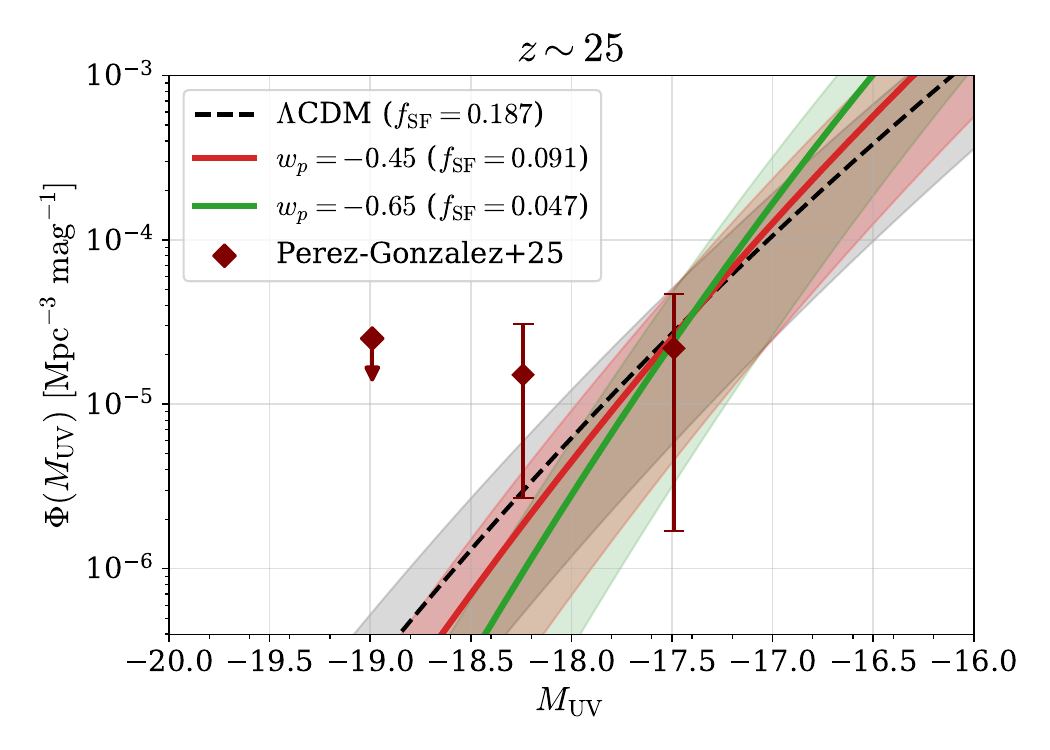}
    \caption{UV LF at $z\sim25$ for Set II with $f_X=f_X^{68}=0.00155$. While the $w_p=-0.45$ model remains consistent with the observed UV luminosity, the $w_p=-0.65$ model (the most negative value allowed by MW satellites for $f_X^{68}$) underpredicts the galaxy abundance around $M_\rm{UV}\simeq-18.3$, with its $1\sigma$ band failing to overlap the corresponding error bar.}
    \label{fig:z25_comp}
\end{figure}

Although a wider range of values, namely $-1<w_p<0$ for Set I and $-0.65\lesssim w_p<0$ for Set II are allowed by Milky Way satellite counts, Lyman-$\a$ forest, and strong lensing constraints, comparison with the observed UV LF provides an additional restriction on the parameter space. In particular, Set II with $w_p\lesssim-0.45$ underpredicts the UV luminosity of the brightest photometric galaxy candidate (with $M_\rm{UV}\simeq-18.3$) at $z\sim25$, while remaining consistent with the UV LF measurements at other redshifts considered in this work. We show this in Fig.~\ref{fig:z25_comp}. This, however, is not in contradiction with the enhanced HMF over the entire halo mass range, since the star-formation efficiency is refitted for each model. So an enhanced HMF does not necessarily translate into an enhanced UV LF at fixed $M_\rm{UV}$. The larger halo abundance allows the observed high-redshift galaxy counts to be reproduced with a smaller $f_\rm{SF}$. For sufficiently strong enhancement, as is the case at $z\sim25$, the best-fit $f_\rm{SF}$ is driven to very small values (as compared to that for $\La$CDM), such that the suppression at the bright end becomes too strong and the model begins to underpredict the UV luminosity.
Since the present $z\sim25$ UV LF is inferred from a small sample of galaxies without spectroscopic confirmation, the significance of this mismatch should be regarded as provisional~\cite{perezgonzalez2025risegalacticempireluminosity}. Nevertheless, we choose to respect this bound, thereby ensuring a conservative estimate for the required $f_\rm{SF}$.

Restricting the analysis to this UV LF-allowed region, \ie $-0.99\leq w_p\leq-0.33$ for Set I and $-0.45\leq w_p\leq-0.33$ for Set II, where the upper limit $w_p=-1/3$ is chosen as the boundary value compatible with an EDE-like interpretation of the exotic component, limits the enhancement of early halo formation relative to the most extreme model (\ie $w_p\approx-0.65$ for Set II). Nevertheless, the modified cosmological models consistently require lower star-formation efficiencies than $\Lambda$CDM to reproduce the observed UV luminosity. For the spectroscopically confirmed samples at $z\sim12$ and $14$, the reduction is modest, whereas for the higher-redshift photometric samples ($z\sim17$ and $25$), it is substantially larger. In particular, the most enhanced UV LF-compatible model (Set II; $w_p=-0.45$) requires a star-formation efficiency that is lower than the $\La$CDM value by a factor of approximately $1.4$ at $z\sim17$, increasing to about a factor of $2$ at $z\sim25$, a conclusion we show to be robust against the choice of HMF in Appendix~\ref{sec:YungHMF}. Although these photometric estimates are subject to larger systematic uncertainties, the overall trend suggests that enhanced early structure formation can partially alleviate the need for unusually high star-formation efficiencies within the standard cosmological model ($f_\rm{SF}\simeq20-65\%$ for the photometric candidates~\cite{Yung_2025}). For completeness, we also report the corresponding results for the 
$95\%$ C.L. upper limits on $f_X$ in Table~\ref{tab:fsf_bestfit}. These should be regarded as limiting cases, illustrating the maximal reduction in $f_\rm{SF}$ permitted by the cosmological constraints. Our model is not statistically favored over $\La$CDM; rather, its main phenomenological advantage is that the observed high-redshift galaxy abundances can be accommodated with a reduced star-formation efficiency.

\section{Discussion and Conclusions}
\label{sec:conclusion}

In this work, we have investigated the impact of a subdominant GDM component, undergoing a transient EDE-like phase, on the formation of early cosmic structures during the JWST era. Although the exotic component constitutes only a sub-percent of the total DM density, its temporary negative EOS parameter gives rise to a finite interval of negative effective sound speed squared, leading to instability-driven growth of density perturbations over a restricted range of comoving scales. 
Using Planck, DESI DR2, and Pantheon+ data, we derived cosmological constraints on the model parameters, and subsequently restricted the viable parameter space by requiring consistency with independent constraints from Milky Way satellite counts, Lyman-$\a$ forest measurements, and strong-lensing observations. Within this allowed region, the transient enhancement of perturbation growth leads to a significant increase in the abundance of DM halos at high redshifts while leaving the large-scale matter power spectrum essentially unchanged. Consequently, the model consistently requires lower star-formation efficiencies than the standard $\Lambda$CDM cosmology in order to reproduce the observed ultraviolet luminosity functions. The reduction is modest for the spectroscopically confirmed galaxies at $z\sim12$--14 but becomes larger for the highest-redshift photometric samples, reaching approximately a factor of two for the most enhanced UV LF-compatible models.


Unlike conventional EDE scenarios (see, for example, the review~\cite{poulin2023upsdownsearlydark} and a recent paper in the JWST context, Ref. \cite{du2026resolvinghubbletensionearly}), whose principal effect is to modify the background expansion history, the mechanism proposed here operates primarily through a transient enhancement of the growth of density perturbations. Since only a small fraction of the DM participates in the transient dynamics, the successful large-scale predictions of the standard $\Lambda$CDM cosmology are largely preserved while the abundance of early DM halos is selectively enhanced over the scales relevant for JWST observations. The present scenario involving non-standard dark sector dynamics should therefore be regarded as complementary to other cosmological proposals based on modified primordial fluctuations or alternative expansion histories, as well as to astrophysical explanations involving enhanced star-formation or AGN activity.

In addition to adopting the GDM framework as an effective description of transient dark-sector dynamics, we have assumed that the galaxy–halo connection and halo accretion histories are adequately described by the standard $\La$CDM calibration, so that the dominant impact of the model enters through the modified HMF. Future hydrodynamical simulations incorporating transient dark-sector dynamics will be required to determine whether changes in halo assembly histories, baryonic feedback, or star-formation physics further modify the predicted UV LFs. Additionally, in this preliminary work, we have kept the astrophysics fixed and have not incorporated halo mass dependence in $f_\rm{SF}$. A more comprehensive analysis, which we keep for future work, would involve using the publicly available package \texttt{GALLUMI}~\cite{Sabti_2022} to marginalize over selected astrophysical nuisance parameters for fixed benchmark values of $w_p$, thereby reconstructing the allowed $f_\rm{SF}(M_h,z)$ consistent with the JWST UV LF. 

Overall, our results demonstrate that a transient EDE-like phase in a subdominant DM component provides a viable and observationally consistent mechanism for enhancing early structure formation. More broadly, this work illustrates that transient departures from standard CDM dynamics can leave observable imprints on galaxy formation during cosmic dawn while remaining compatible with current cosmological constraints, thereby providing a new phenomenological framework for interpreting present and future observations of the high-redshift Universe.

\begin{acknowledgments}
We would like to thank Ethan Nadler for useful discussions. AM would like to thank the Leinweber Institute for Theoretical Physics for supporting his sabbatical visit at the University of Michigan, Ann Arbor.
\end{acknowledgments}

\appendix

\section{A Possible Microscopic Interpretation}
\label{sec:micro}
\setcounter{figure}{0}
\renewcommand{\thefigure}{A\arabic{figure}}
\setcounter{table}{0}
\renewcommand{\thetable}{A-\Roman{table}}

The phenomenological framework developed in this work is intentionally agnostic about the underlying particle physics. Our primary objective is to investigate the cosmological consequences of transient dark sector dynamics rather than to construct a complete microscopic theory. Nevertheless, it is useful to ask whether the characteristic sequence
\begin{align}\label{seq}
w_X\simeq 0
\quad\longrightarrow\quad
w_X<0
\quad\longrightarrow\quad
w_X\simeq 0
\end{align}
can arise from an underlying field-theoretic construction.

A possible realization is a coupled dark sector consisting of a non-relativistic fermion $\psi$ and an axion-like scalar $\phi$, described schematically by
\begin{equation}
\mathcal{L}_X=
-\frac{1}{2}(\partial\phi)^2-V(\phi)
+\bar\psi\left(i\slashed{\partial}-m_\psi(\phi)\right)\psi .
\end{equation}
The fermionic component remains non-relativistic throughout the epoch of interest and hence has negligible pressure, $p_\psi\simeq0$. The scalar may evolve in a multi-periodic potential generated by several non-perturbative contributions, or, more generally, $\phi$ may represent an effective light direction obtained after integrating out heavier fields in a multi-axion potential. Such potentials can possess multiple metastable minima and non-trivial interpolating regions, as encountered in axion-landscape and string-inspired constructions~\cite{Arvanitaki_2010,Bachlechner_2018}.

The background quantities relevant to our phenomenological fluid are those of the combined scalar--fermion sector,
\begin{align}
\rho_X=\rho_\psi+\frac{1}{2}\dot\phi^2+V(\phi),\quad 
p_X\simeq\frac{1}{2}\dot\phi^2-V(\phi),
\end{align}
where $\rho_\psi\simeq m_\psi(\phi)n_\psi$ ($n_\psi$ being the fermion number density) and hence
\begin{equation}
w_X=
\frac{\frac{1}{2}\dot\phi^2-V(\phi)}
{\rho_\psi+\frac{1}{2}\dot\phi^2+V(\phi)}.
\end{equation}
Although energy may be exchanged internally between $\phi$ and $\psi$ through the $\phi$ dependence of the fermion mass, the total scalar--fermion sector can remain separately conserved~\cite{Amendola_2000}.

At early times, suppose that the axion undergoes rapid coherent oscillations about an approximately quadratic local minimum, yielding $\l<w_\phi\r>\simeq0$. Since the fermion is also pressureless, the combined sector behaves as matter, $\langle w_X\rangle\simeq0$. During radiation domination, a non-perturbative transition, such as tunneling between metastable regions of the axion potential, may place $\phi$ away from its final minimum. If it emerges in a sufficiently shallow region, its evolution can become temporarily potential dominated. The fermion remains pressureless during this process, while the scalar supplies a negative pressure. Consequently, the equation of state of the combined fluid becomes
\begin{align}
w_X\simeq
-\frac{V(\phi)}
{\rho_\psi+V(\phi)}<0 .
\end{align}
Thus the scalar itself may be close to a vacuum-energy-like state, $w_\phi\simeq-1$, while the presence of the pressureless fermionic component naturally yields a less negative effective $w_X$. As the scalar subsequently settles into another approximately quadratic minimum, coherent oscillations resume, and the combined equation of state returns to $\l<w_X\r>\to0$.

This interpretation has close structural analogy with the sterile-neutrino--pseudoscalar fluid of Ref.~\cite{sharma2026recoupleddarkradiationreconciling}. The combined fluid initially has $w=1/3$, departs from this value when the sterile neutrinos become non-relativistic, and returns to $1/3$ after their annihilation into relativistic pseudoscalars. In the present case, the analogous microscopic event is instead the transition of the axion between different regions of its potential, producing the sequence (\ref{seq}). The Gaussian profile in Eq.~(\ref{eos}) can therefore be regarded as a smooth coarse-grained parametrization of this transient rearrangement of the scalar--fermion dark sector, rather than as the exact equation of state generated by a particular axion potential.

For the perturbation sector, the scale $\kappa$ cannot be identified with the mass of the axion-like field. In the parameter range relevant for our analysis, $\kappa<H$ during the transient epoch, whereas interpreting the Lorentzian factor in Eq.~(\ref{cs2}) as arising from the propagator of $\phi$ would require a quasi-static or integrated-out scalar degree of freedom, typically with $m_\phi\gtrsim H$. We should therefore regard $\kappa$ as an effective response scale of the coupled scalar–fermion medium rather than as a fundamental particle mass.

Within the coupled scalar--fermion picture, the pressure perturbation is carried predominantly by the scalar sector, whereas the total density perturbation receives contributions from both the scalar and the pressureless fermions. The system therefore possesses an entropy perturbation which allows the effective pressure response of the combined fluid to differ from the purely adiabatic one. Defining the non-adiabatic pressure perturbation as $\delta p_{X,\rm{nad}}\equiv\d p_X-c_a^2\d\rho_X$ ~\cite{Malik_2005,Ballesteros_2010}, and using Eq.~(\ref{dpX}), one obtains
\begin{align}
    \d p_{X,\rm{nad}}=(c_s^2-c_a^2)\l(\d\rho_X-\dot{\bar\rho}_X\frac{\theta_X}{k^2}\r)=(c_s^2-c_a^2)\d\rho_X^\rm{(rf)}.
\end{align}
Hence the GDM rest-frame sound speed squared may be equivalently written as
\begin{align}
    c_s^2=c_a^2+\frac{\d p_{X,\rm{nad}}}{\d\rho_X^\rm{(rf)}}.
\end{align}
Our parametrization in Eq.~(\ref{cs2}) therefore corresponds to assuming that the non-adiabatic contribution has a scale dependence $k^2/(k^2+a^2\kappa^2)$, with the proportionality constant given by $-c_a^2$.
The Lorentzian form may be viewed as the simplest one-scale interpolation between a regime in which the combined fluid follows the adiabatic response and a regime in which the relative scalar--fermion dynamics increasingly reduce the effective pressure perturbation. 

\section{Dependence on the Effective Scale $\kappa_0$}
\label{sec:kappa0}
\setcounter{figure}{0}
\renewcommand{\thefigure}{B\arabic{figure}}
\setcounter{table}{0}
\renewcommand{\thetable}{B-\Roman{table}}
In this appendix, we examine the dependence of our results on the effective scale $\kappa_0$ more systematically. In Fig.~\ref{fig:vary_kappa0} we plot the ratio of the modified matter power spectrum to the $\La$CDM matter power spectrum for other values of $\kappa_0$, along with constraints from Milky Way satellite galaxy counts~\cite{10.1093/mnras/stz2310} and strong lensing measurements~\cite{10.1093/mnras/stac670}, indicated by blue and green shaded regions, respectively. In order to isolate the impact of varying $\{w_p,\kappa_0\}$ we fix the standard cosmological parameters to their $\La$CDM best-fit values and set $f_X=0.1\%$. From Fig.~\ref{fig:vary_kappa0} (and also Fig.~\ref{fig:kappa}), it is evident that the minimum value of $w_p$ consistent with the displayed bounds over the range $k\sim1-100~h/\rm{Mpc}$ increases if we increase the value of $\kappa_0$. We consider $\kappa_0=5.5,\,8.5,\,12.0~h/\rm{Mpc}$, along with their corresponding minimum values of $w_p$ (\viz $w_p=-0.99$, $w_p=-0.38$, $w_p=-0.18$, respectively) compatible with the above mentioned constraints, as supplementary test cases. We also consider another set of $w_p$ values ($w_p'=-0.65$, $w_p'=-0.25$, $w_p'=-0.10$, respectively), denoted by a prime. These cases are not subject to separate MCMC analyses, but are introduced solely to illustrate how alternative choices of $\kappa_0$ affect the inferred $f_\rm{SF}$.
\begin{figure}[htbp]
    \centering
    \includegraphics[width=\linewidth]{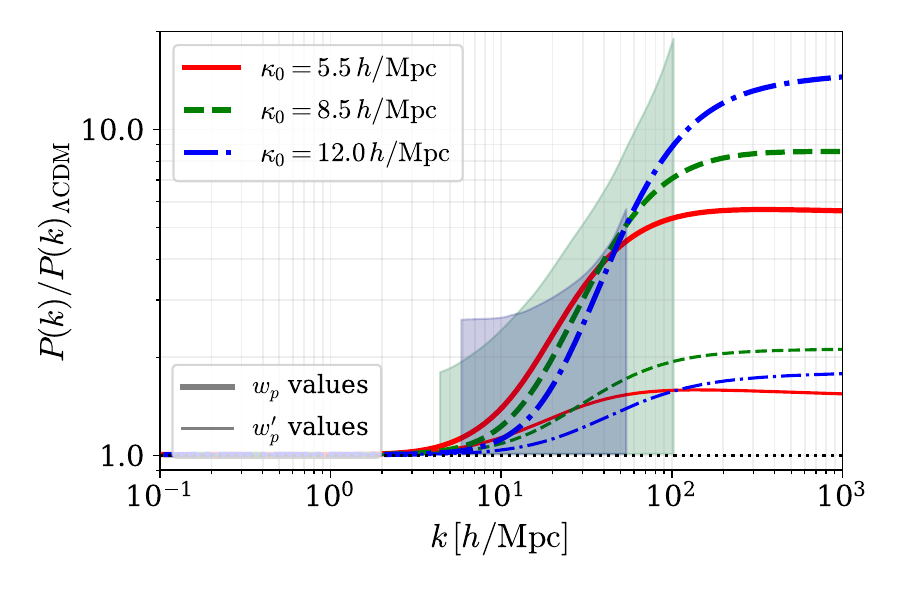}
    \caption{Ratio of the modified and the $\La$CDM matter power spectra for some values of $\kappa_0$ other than our benchmark values (and with $f_X=0.1\%$). The blue and green shaded regions correspond to the constraints from MW dwarf galaxies and strong lensing, respectively.}
    \label{fig:vary_kappa0}
\end{figure}
\begin{figure}[htbp]
    \centering
    \begin{subfigure}{0.48\textwidth}
        \centering
        \includegraphics[width=\textwidth]{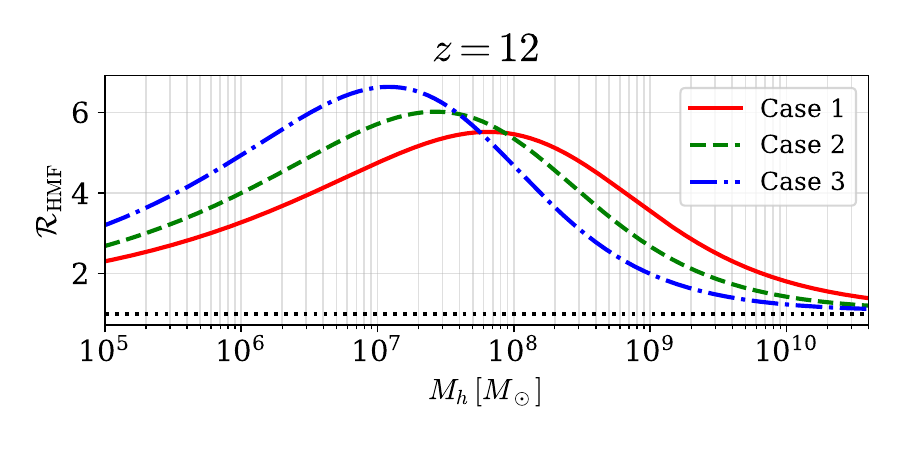}
    \end{subfigure}
    \hfill 
    \begin{subfigure}{0.48\textwidth}
        \centering
        \includegraphics[width=\textwidth]{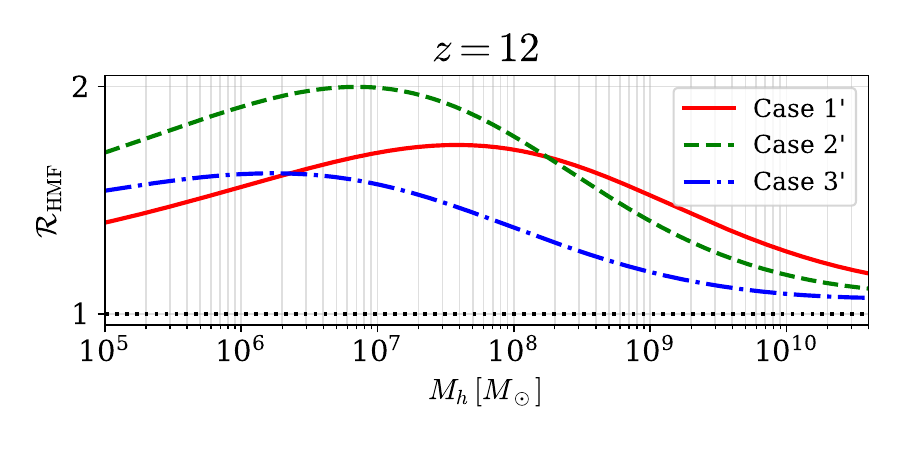}
    \end{subfigure}
    \hfill
    \begin{subfigure}{0.48\textwidth}
        \centering
        \includegraphics[width=\textwidth]{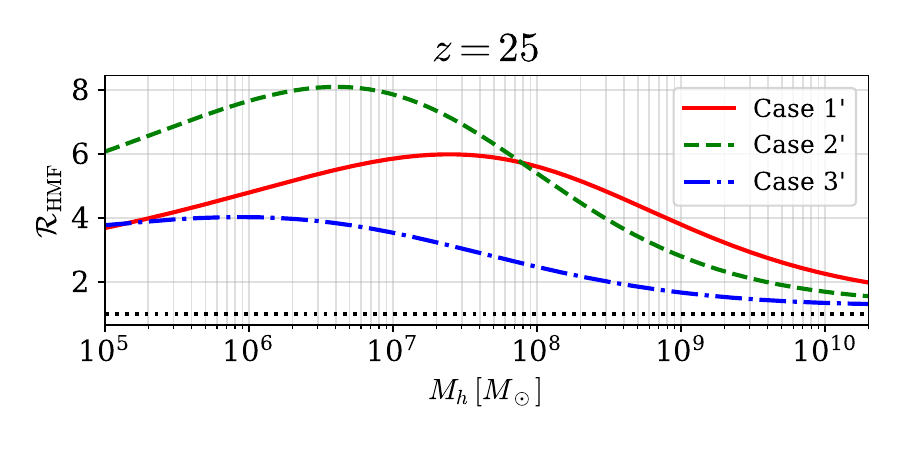}
    \end{subfigure}
    \caption{Ratio of the HMFs for the supplementary test cases. Unprimed cases correspond to the minimum $w_p$ values compatible with MW satellite galaxies/strong lensing measurements whereas primed cases correspond to minimum $w_p$ values that are additionally compatible with $z\sim25$ UV LF constraints.}
    \label{fig:STC_hmf}
\end{figure}

We label the supplementary test cases as Case 1: $\{\kappa_0=5.5~h/\rm{Mpc},w_p=-0.99\}$, Case 2: $\{\kappa_0=8.5~h/\rm{Mpc},w_p=-0.38\}$, and Case 3: $\{\kappa_0=12.0~h/\rm{Mpc},w_p=-0.18\}$, all with $f_X=0.1\%$. We show the ratios of the HMFs, $\mathcal R_\rm{HMF}$ at $z=12$, in the top panel of Fig.~\ref{fig:STC_hmf}. As mentioned previously, consistency of the predicted UV LF at $z\sim25$ shrinks the allowed $w_p$ range: $w_p=-0.99\to-0.65$ (Case $1'$), $w_p=-0.38\to-0.25$ (Case $2'$), and $w_p=-0.18\to-0.10$ (Case $3'$). The corresponding $\mathcal R_\rm{HMF}$ at $z=12$ and $z=25$ are shown in the middle and bottom panels of Fig.~\ref{fig:STC_hmf}. The resulting best-fit star-formation efficiencies are given in Table~\ref{tab:fsf_stc}.
\begin{table}[ht]
    \centering
    \setlength{\tabcolsep}{3.5pt}
    \begin{tabular}{llccc}
        \hline\hline
        $z$ & $w_p$ constrained by & Case 1 ($1'$) & Case 2 ($2'$) & Case 3 ($3'$) \\
        \hline
        12 & MW sat.+lensing & 0.0496 & 0.0520 & 0.0533 \\
        12 & + high-$z$ UV LF & 0.0525 & 0.0535 & 0.0543 \\
        25 & + high-$z$ UV LF & 0.1126 & 0.1181 & 0.1456 \\
        \hline\hline
    \end{tabular}
    \caption{Derived star formation efficiencies for different redshifts and constraint scenarios for the supplementary test cases.}
    \label{tab:fsf_stc}
\end{table}
One can make the following observations. 
\begin{itemize}
    \item Firstly, the $\mathcal R_\rm{HMF}$ peak amplitude and position depend on both $w_p$ and $\kappa_0$. While $\kappa_0$ predominantly affects the peak location, $w_p$ primarily controls the strength of the enhancement. A comparison of the three curves in any panel of Fig.~\ref{fig:STC_hmf} indicates that the peaks shift horizontally towards smaller halo masses with increasing $\kappa_0$, \ie $\kappa_0$ determines the characteristic halo mass scale (however, the quantitative relation is not obvious). Next, comparing the top and middle panels, one can isolate the effect of $w_p$ on $\mathcal R_\rm{HMF}$: more negative $w_p$ leads to more enhancement, as expected. Also, increasing $w_p$ (making it less negative) has an opposite effect to increasing $\kappa_0$. Finally, comparing the middle and bottom panels shows that the enhancement is stronger at higher redshifts, as we have already observed (in Fig.~\ref{fig:HMF}).
    \item Secondly, while $f_\rm{SF}$ is smaller than that of $\La$CDM for all three cases, Case 1 (Case $1'$) shows the maximum improvement over $\La$CDM while the improvement for Case 3 (Case $3'$) is the least. This is in spite of the different hierarchy of the peak heights among the corresponding curves in the top and middle panels of Fig.~\ref{fig:STC_hmf}. The reason is that the inferred $f_\rm{SF}$ reflects not only the magnitude of the HMF enhancement but also the characteristic halo mass scale where $\mathcal R_\rm{HMF}$ peaks. This occurs because the enhancement in Case 1 extends to comparatively larger halo masses, which overlap more efficiently with the halo mass range relevant for the UV-bright JWST galaxies at the corresponding redshifts. This illustrates that the relevant quantity for the UV LF is the enhancement of the halo abundance over the mass range populated by the observed galaxies/galaxy candidates, rather than the maximum value of $\mathcal R_\rm{HMF}$ itself.
\end{itemize}
A value of $\kappa_0\gtrsim10~h/\rm{Mpc}$ would require a less negative $w_p$ value to be compatible with observations over the range $k\sim1-100~h/\rm{Mpc}$ as well as with high-$z$ (photometric) UV LF. This in turn would worsen the improvement in $f_\rm{SF}$, as compared to the case with $6\lesssim\kappa_0\lesssim10~h/\rm{Mpc}$. Our benchmark value $\kappa_0=3.6~h/\rm{Mpc}$ shows mild improvement over $\La$CDM when paired with the extreme value of $w_p$, \ie $w_p\approx-0.99$, while the other benchmark value, $\kappa_0=6.4~h/\rm{Mpc}$, exemplifies a case yielding a larger allowed enhancement with a moderate value of $w_p$.

\section{Robustness to the choice of HMF}
\label{sec:YungHMF}
\setcounter{figure}{0}
\renewcommand{\thefigure}{C\arabic{figure}}
\setcounter{table}{0}
\renewcommand{\thetable}{C-\Roman{table}}
\begin{figure}[htbp]
    \centering
    \includegraphics[width=\linewidth]{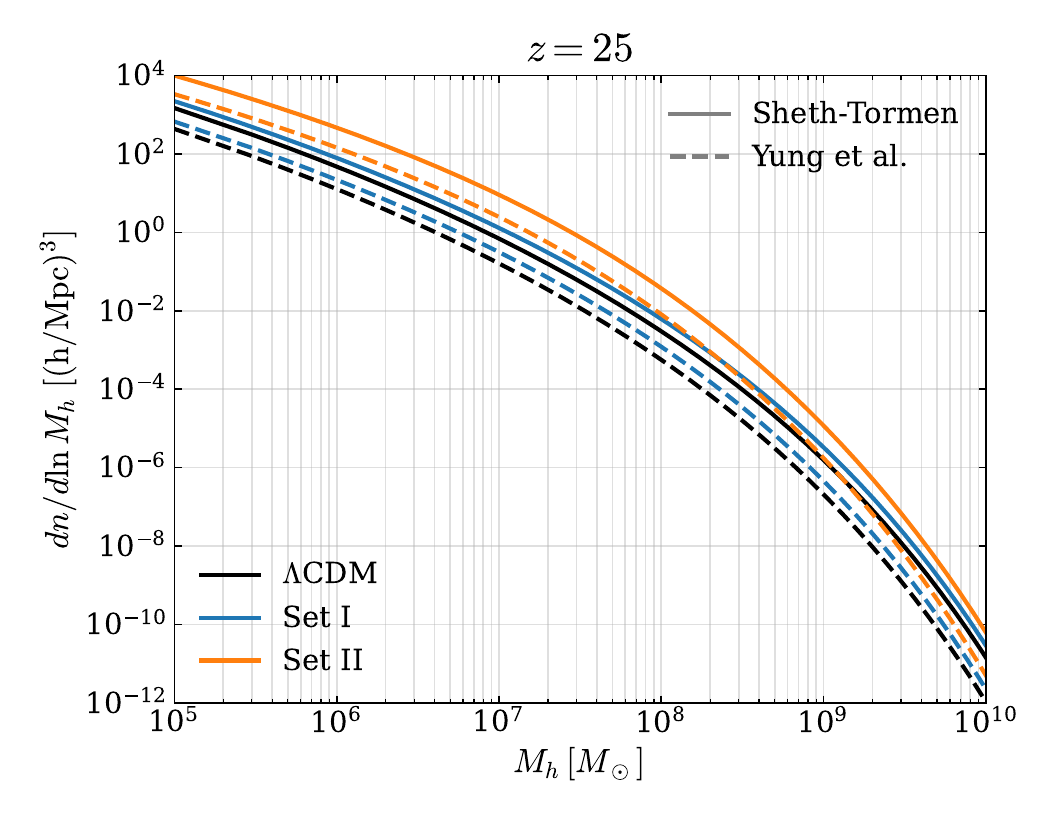}
    \caption{Comparison of the standard ST HMF with the $N$-body simulations-calibrated Yung et al. HMF.}
    \label{fig:HMF_comp}
\end{figure}
Throughout the main analysis, we adopted the Sheth-Tormen (ST) HMF as our fiducial prescription. While adopting the ST form is standard practice for predicting the abundance of collapsed halos, its extrapolation to the very high-redshift regime probed by JWST is subject to increasing theoretical uncertainty. Recent high-resolution $N$-body simulations have provided improved calibrations of the HMF at $z\gtrsim 10$, including the GUREFT-based fitting function of Yung et al.~\cite{Yung_2025}, which was specifically constructed to describe halo abundances at very high redshift. An important caveat is that the Yung et al. calibration is obtained from $\La$CDM $N$-body simulations, as mentioned in footnote~\ref{footnote:yung}. Here we assume their multiplicity function remains applicable when evaluated with our modified linear variance. This assumption is not guaranteed in a general non-standard DM cosmology. In the present setup, however, the exotic component constitutes only a sub-percent fraction of the DM and the primary remnant of the transient epoch is therefore a modified linear matter power spectrum. 

\begin{table}[b]
    \centering
    \setlength{\tabcolsep}{4pt}
    \begin{tabular}{c|c|c|c}
        \hline\hline
        \multirow{3}{*}{$z$}
        & \multirow{3}{*}{$\Lambda$CDM}
        & Set I
        & Set II
        \\
        \cline{3-4}
        &
        &
        $f_X^{68}=0.00161$
        &
        $f_X^{68}=0.00155$
        \\
        \cline{3-4}
        &
        &
        $w_p=-0.99$
        &
        $w_p=-0.45$
        \\
        \hline
        17
        & $0.074_{-0.020}^{+0.015}$
        & $0.063_{-0.017}^{+0.012}$
        & $0.048_{-0.013}^{+0.009}$
        \\
        25
        & $0.332_{-0.130}^{+0.079}$
        & $0.256_{-0.099}^{+0.060}$
        & $0.145_{-0.054}^{+0.031}$
        \\
        \hline\hline
    \end{tabular}
    \caption{best-fit star-formation efficiencies obtained using the Yung et al. HMF.}
    \label{tab:yung_fsf}
\end{table}

Yung et al. adopted the following fitting function (cf.~Eq.~(\ref{st})):
\begin{align}
    f(\sigma)=A\l[\l(\frac{\sigma}{b}\r)^{-a}+1\r]e^{-c/\sigma^2},
\end{align}
where $\chi_i=A,a,b$, and $c$ are redshift dependent and are parameterized as $\chi_i(z)=\chi_{0,i}+\chi_{1,i}z+\chi_{2,i}z^2$, with the coefficients given in Table~1 of Ref.~\cite{Yung_2025}. We repeat the UV LF analyses at $z\sim17$ and $z\sim25$, where the enhancement of the halo abundance and its impact on the inferred star-formation efficiency are largest. We compare the Yung et al. HMF with the ST HMF in Fig.~\ref{fig:HMF_comp}. The corresponding best-fit values of $f_\rm{SF}$ are summarized in Table~\ref{tab:yung_fsf}. Although the absolute values of the inferred star-formation efficiencies shift relative to those obtained using the ST prescription, the qualitative hierarchy between the cosmologies remains unchanged. In particular, the two benchmark cases continue to require lower star-formation efficiencies than $\La$CDM, with Set II exhibiting reductions by approximately a factor of $1.4$ at $z\sim17$ and $2$ at $z\sim25$, essentially the same relative improvement found with the fiducial ST HMF.

This comparison indicates that the principal conclusion of our analysis---namely, the quantitative reduction in $f_\rm{SF}$ with respect to $\La$CDM owing to the transient dark sector dynamics---is not specific to the choice of the Sheth-Tormen HMF. A fully self-consistent determination of the halo abundance in the present scenario would ultimately require dedicated $N$-body simulations incorporating the modified dark-sector dynamics.

\bibliographystyle{apsrev4-2}
\bibliography{references}

\end{document}